\documentclass[10pt,twocolumn,twoside,journal]{IEEEtran}
\usepackage{amsmath}
\usepackage{times}
\usepackage{amsbsy,color}
\usepackage{latexsym}
\usepackage{amssymb}
\usepackage{graphicx}
\usepackage{subfigure}
\usepackage{amsfonts}
\usepackage{epsfig,subfigure}
\allowdisplaybreaks[4] %
\begin{document}

\title{\huge Dynamic Topology Adaptation Based on Adaptive Link Selection Algorithms for Distributed Estimation}
\author{Songcen~Xu*,~
        Rodrigo C. de~Lamare,~\IEEEmembership{Senior Member,~IEEE,}
        and~H. Vincent~Poor,~\IEEEmembership{Fellow,~IEEE}  \vspace{-1em}% <-this % stops a space
\thanks{S. Xu* is with the Communications Research Group, Department of Electronics, University of York, YO10 5DD York, U.K.
(e-mail: songcen.xu@york.ac.uk).}% <-this % stops a space
\thanks{R. C. de Lamare is with CETUC / PUC-Rio, Brazil and Department of Electronics, University of York, U.K.
(e-mail: rodrigo.delamare@york.ac.uk).}% <-this % stops a space
\thanks{H. V. Poor is with the Department of Electrical Engineering, Princeton University,
Princeton NJ 08544 USA (e-mail: poor@princeton.edu).}% <-this % stops a space
\thanks{Part of this work has been presented at the 2013 IEEE International Conference on Acoustics, Speech, and Signal Processing, Vancouver, Canada and 2013 IEEE International Workshop on Computational Advances in Multi-Sensor Adaptive Processing, Saint Martin. EDICS: NET-DISP, NET-ADEG, NET-GRPH, NET-SPRS}}

% The paper headers
%\markboth{Journal of \LaTeX\ Class Files,~Vol.~11, No.~4, December~2012}%
%{Shell \MakeLowercase{\textit{et al.}}: Bare Demo of IEEEtran.cls for Journals}

\maketitle

\begin{abstract}
This paper presents adaptive link selection
algorithms for distributed estimation and considers their
application to wireless sensor networks and smart grids. In
particular, exhaustive search--based
least--mean--squares(LMS)/recursive least squares(RLS) link
selection algorithms and sparsity--inspired LMS/RLS link selection
algorithms that can exploit the topology of networks with
poor--quality links are considered. The proposed link selection
algorithms are then analyzed in terms of their stability,
steady--state and tracking performance, and computational
complexity. In comparison with existing centralized or distributed
estimation strategies, key features of the proposed algorithms are:
1) more accurate estimates and faster convergence speed can be
obtained; and 2) the network is equipped with the ability of link
selection that can circumvent link failures and improve the
estimation performance. The performance of the proposed algorithms
for distributed estimation is illustrated via simulations in
applications of wireless sensor networks and smart grids.
\end{abstract}


\IEEEpeerreviewmaketitle

\vspace{-1.2em}
\section{Introduction}
\IEEEPARstart{D}{istributed} signal processing algorithms have
become a key approach for statistical inference in wireless networks
and applications such as wireless sensor networks and smart grids
\cite{Lopes1,Lopes2,Chen,Xie}. It is well known that distributed
processing techniques deal with the extraction of information from
data collected at nodes that are distributed over a geographic area
\cite{Lopes1}. In this context, for each specific node, a set of
neighbor nodes collect their local information and transmit the
estimates to a specific node. Then, each specific node combines the
collected information together with its local estimate to generate
an improved estimate. \vspace{-0.5em}

\subsection{Prior and Related Work}
Several works in the literature have proposed strategies for distributed
processing which include incremental \cite{Lopes1,Bertsekas,Bertsekas2,Nowak},
diffusion \cite{Lopes2,Cattivelli3}, sparsity--aware
\cite{Chen,Lorenzo,dce,dts} and consensus--based strategies \cite{Xie}. With
the incremental strategy, the processing follows a Hamiltonian cycle, i.e., the
information flows through these nodes in one direction, which means each node
passes the information to its adjacent node in a uniform direction. However, in
order to determine a cyclic path that covers all nodes, this method needs to
solve an NP--hard problem. In addition, when any of the nodes fails, data
communication through the cycle is interrupted and the distributed processing
breaks down \cite{Lopes1}.

In distributed diffusion strategies \cite{Lopes2,Lorenzo}, the
neighbors for each node are fixed and the combining coefficients are
calculated after the network topology is deployed and starts its
operation. One disadvantage of this approach is that the estimation
procedure may be affected by poorly performing links. More
specifically, the fixed neighbors and the pre--calculated combining
coefficients may not provide an optimized estimation performance for
each specified node because there are links that are more severely
affected by noise or fading. Moreover, when the number of neighbor
nodes is large, each node requires a large bandwidth and transmit
power. Prior work on topology design and adjustment techniques
includes the studies in \cite{Lopes11,Bilal} and \cite{Wimalajeewa},
which are not dynamic in the sense that they cannot track changes in
the network and mitigate the effects of poor links. \vspace{-0.5em}

\subsection{Contributions}
This paper proposes and studies adaptive link
selection algorithms for distributed estimation problems.
Specifically, we develop adaptive link selection algorithms that can
exploit the knowledge of poor links by selecting a subset of data
from neighbor nodes.  The first approach consists of exhaustive
search--based least--mean--squares(LMS)/recursive least squares(RLS)
link selection (ES--LMS/ES--RLS) algorithms, whereas the second
technique is based on sparsity--inspired LMS/RLS link selection
(SI--LMS/SI--RLS) algorithms. With both approaches, distributed
processing can be divided into two steps. The first step is called
the adaptation step, in which each node employs LMS or RLS to
perform the adaptation through its local information. Following the
adaptation step, each node will combine its collected estimates from
its neighbors and local estimate, through the proposed adaptive link
selection algorithms. The proposed algorithms result in improved
estimation performance in terms of the mean--square error (MSE)
associated with the estimates. In contrast to previously reported
techniques, a key feature of the proposed algorithms is that the
combination step involves only a subset of the data associated with
the best performing links.

In the ES--LMS and ES--RLS algorithms, we consider
all possible combinations for each node with its neighbors and
choose the combination associated with the smallest MSE value. In
the SI--LMS and SI--RLS algorithms, we incorporate a reweighted zero
attraction (RZA) strategy into the adaptive link selection
algorithms. The RZA approach is often employed in applications
dealing with sparse systems in such a way that it shrinks the small
values in the parameter vector to zero, which results in better
convergence and steady--state performance. Unlike prior work with
sparsity--aware algorithms \cite{Chen,Rcdl1,Rcdl2,Guo}, the proposed
SI--LMS and SI--RLS algorithms exploit the possible sparsity of the
MSE values associated with each of the links in a different way. In
contrast to existing methods that shrink the signal samples to zero,
SI--LMS and SI--RLS shrink to zero the links that have poor
performance or high MSE values. By using the
SI--LMS and SI--RLS algorithms, data associated with unsatisfactory
performance will be discarded, which means the effective network
topology used in the estimation procedure will change as well.
Although the physical topology is not changed by the proposed
algorithms, the choice of the data coming from the neighbor nodes
for each node is dynamic, leads to the change of combination weights
and results in improved performance. We also remark that the
topology could be altered with the aid of the proposed algorithms
and a feedback channel which could inform the nodes whether they
should be switched off or not. The proposed algorithms are
considered for wireless sensor networks and also as a tool for
distributed state estimation that could be used in smart grids.

In summary, the main contributions of this paper are:
\begin{itemize}
\item We present adaptive link selection algorithms for distributed
estimation that are able to achieve significantly better
performance than existing algorithms.
\item We devise distributed LMS and RLS algorithms with link
selection capabilities to perform distributed estimation.
\item We analyze the MSE convergence and tracking performance of
the proposed algorithms and their computational complexities and we derive
analytical formulas to predict their MSE performance.
\item A simulation study of the proposed and existing
distributed estimation algorithms is conducted along with
applications in wireless sensor networks and smart grids.
\end{itemize}

This paper is organized as follows. Section II describes the system
model and the problem statement. In Section III, the proposed link
selection algorithms are introduced. We analyze the proposed
algorithms in terms of their stability, steady--state and tracking
performance, and computational complexity in Section IV. The
numerical simulation results are provided in Section V. Finally, we
conclude the paper in Section VI.

Notation: We use boldface upper case letters to denote matrices and
boldface lower case letters to denote vectors. We use $(\cdot)^T$
and $(\cdot)^{-1}$ denote the transpose and inverse operators
respectively, $(\cdot)^H$ for conjugate transposition and
$(\cdot)^*$ for complex conjugate.
%\vspace{-0.25em}

\section{System Model and Problem Statement}

\vspace{-0.2em}
\begin{figure}[!htb]
\begin{center}
\def\epsfsize#1#2{0.325\columnwidth}
\epsfbox{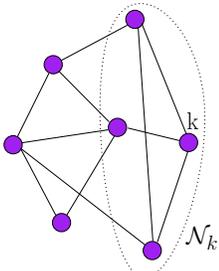} \vskip -10pt\caption{\footnotesize
Network topology with $N$ nodes} \vskip -15pt \label{fig1}
\end{center}
\end{figure}

We consider a set of $N$ nodes, which have limited
processing capabilities, distributed over a given geographical area
as depicted in Fig. \ref{fig1}. The nodes are connected and form a
network, which is assumed to be partially connected because nodes
can exchange information only with neighbors determined by the
connectivity topology. We call a network with this property a
partially connected network whereas a fully connected network means
that data broadcast by a node can be captured by all other nodes in
the network in one hop \cite{Bertrand}. We can think of this network
as a wireless network, but our analysis also applies to wired
networks such as power grids. In our work, in order to perform link
selection strategies, we assume that each node has at least two
neighbors.

The aim of the network is to estimate an unknown parameter vector
${\boldsymbol{\omega}}_{0}$, which has length $M$. At every time
instant $i$, each node $k$ takes a scalar measurement $d_k(i)$
according to
\begin{equation}
{d_k(i)} = {\boldsymbol {\omega}}_{0}^H{\boldsymbol x_k(i)}
+{n_k(i)},~~~ i=1,2, \ldots, I ,\label{d_k}
\end{equation}
where ${\boldsymbol x_k(i)}$ is the $M \times 1$ random regression
input signal vector and ${n_k(i)}$ denotes the Gaussian noise at
each node with zero mean and variance $\sigma_{n,k}^2$. This linear
model is able to capture or approximate well many input-output
relations for estimation purposes \cite{Haykin} and we assume $I>M$.
To compute an estimate of ${\boldsymbol{\omega}}$ in a distributed
fashion, we need each node to minimize the MSE cost function
\cite{Lopes2}
\begin{equation}
{J_{\boldsymbol\omega_k(i)}\big({\boldsymbol \omega_k(i)}\big)} =
{\mathbb{E} \big|{ d_k(i)}- {\boldsymbol \omega_k^H(i)}{\boldsymbol
x_k(i)}\big|^2}, \label{cost_function}
\end{equation}
where $\mathbb{E}$ denotes expectation and $\boldsymbol
\omega_k(i)$ is the estimated vector generated by node $k$ at time
instant $i$. Equation (\ref{cost_function}) is also the definition
of the MSE. To solve this problem, diffusion strategies have been
proposed in \cite{Lopes2,Cattivelli3} and \cite{Li}. In these
strategies, the estimate for each node is generated through a fixed
combination strategy given by
\begin{equation}
{\boldsymbol {\omega}}_k(i)= \sum\limits_{l\in \mathcal{N}_k} c_{kl}
\boldsymbol\psi_l(i), \label{comb_strat}
\end{equation}
where $\mathcal{N}_k$ denotes the set of neighbors of node $k$
including node $k$ itself, $c_{kl}\geq0$ is the
combining coefficient and $\boldsymbol\psi_l(i)$ is the local
estimate generated by node $l$ through its local information.

There are many ways to calculate the combining
coefficient $c_{kl}$ which include the Hastings \cite{Zhao1}, the
Metropolis \cite{Xiao}, the Laplacian \cite{Olfati} and the nearest
neighbor \cite{Jadbabaie} rules. In this work, due to its simplicity
and good performance we adopt the Metropolis rule \cite{Xiao} given
by
\begin{equation}
c_{kl}=\left\{\begin{array}{ll}
\frac{1}
{max\{|\mathcal{N}_k|,|\mathcal{N}_l|\}},\ \ $if\  $k\neq l$\ are linked$\\
1 - \sum\limits_{l\in \mathcal{N}_k / k} c_{kl}, \ \ $for\  $k$\ =\ $l$$.
\end{array}
\right. \label{Metropolis}
\end{equation}
where $|\mathcal{N}_k|$ denotes the cardinality of $\mathcal{N}_k$.

The set of coefficients $c_{kl}$ should satisfy \cite{Lopes2}
\begin{equation}
\sum\limits_{l\in \mathcal{N}_k \ \forall k} c_{kl} =1.
\label{condition_sum1}
\end{equation}

For the combination strategy mentioned in (\ref{comb_strat}), the
choice of neighbors for each node is fixed, which results in some
problems and limitations, namely:
\begin{itemize}
\item Some nodes may face high levels of noise or interference, which may lead to inaccurate estimates.
\item When the number of neighbors for each node is high, large communication bandwidth and high transmit power are required.
\item Some nodes may shut down or collapse due to network problems. As a result, local estimates to their neighbors may be affected.
\end{itemize}
Under such circumstances, a performance degradation
is likely to occur when the network cannot discard the contribution
of poorly performing links and their associated data in the
estimation procedure. In the next section, the proposed adaptive
link selection algorithms are presented, which equip a network with
the ability to improve the estimation procedure. In the proposed
scheme, each node is able to dynamically select the data coming from
its neighbors in order to optimize the performance of distributed
estimation techniques. \vspace{-0.5em}

\section{Proposed Adaptive Link Selection Algorithms}

In this section, we present the proposed adaptive link selection
algorithms. The goal of the proposed algorithms is to optimize the
distributed estimation and improve the performance of the network by
dynamically changing the topology. These algorithmic strategies give
the nodes the ability to choose their neighbors based on their MSE
performance. We develop two categories of adaptive link selection
algorithms; the first one is based on an exhaustive search, while
the second is based on a sparsity--inspired relaxation. The details
will be illustrated in the following
subsections. %The proposed can be applied in most distributed
%estimation environments such as wireless sensor networks and smart
%grids.
\vspace{-0.2em}

\subsection{Exhaustive Search--Based LMS/RLS Link Selection }
\vspace{-0.3em}

The proposed ES--LMS and ES--RLS algorithms employ an exhaustive
search to select the links that yield the best performance in terms
of MSE. First, we describe how we define the adaptation step for
these two strategies. In the ES--LMS algorithm, we employ the
adaptation strategy given by
\begin{equation}
{\boldsymbol {\psi}}_k(i)= {\boldsymbol {\omega}}_k(i-1)+{\mu}_k {\boldsymbol x_k(i)}\big[{ d_k(i)}-
{\boldsymbol \omega}_k^H(i-1){\boldsymbol x_k(i)}\big]^*, \label{LMS_adapt}
\end{equation}
where ${\mu}_k$ is the step size for each node. In the ES--RLS
algorithm, we employ the following steps for the adaptation:
\begin{eqnarray}
\boldsymbol\phi^{-1}(i)&=& \lambda^{-1}  \boldsymbol\phi^{-1}(i-1) \notag\\
&& - \dfrac {\lambda ^{-2}\boldsymbol\phi^{-1}(i-1) \boldsymbol x(i) \boldsymbol x^H(i)\boldsymbol\phi^{-1}(i-1)}
{1+\lambda^{-1} \boldsymbol x^H(i) \boldsymbol\phi^{-1}(i-1)\boldsymbol x(i)},
\end{eqnarray}
where $\lambda$ is the forgetting factor. Then, we let
\begin{equation}
\boldsymbol P(i)=\boldsymbol\phi^{-1}(i)
\end{equation}
and
\begin{equation}
\boldsymbol k(i)= \dfrac {\lambda ^{-1}\boldsymbol P(i) \boldsymbol x(i) }
{1+\lambda^{-1} \boldsymbol x^H(i)\boldsymbol P(i)\boldsymbol x(i)}.
\end{equation}
\begin{equation}
{\boldsymbol {\psi}}_k(i)= {\boldsymbol {\omega}}_k(i-1)+ {\boldsymbol k(i)}\big[{ d_k(i)}-
{\boldsymbol \omega}_k^H(i-1){\boldsymbol x_k(i)}\big]^*,\label{ES_RLS1}
\end{equation}
\begin{equation}
\boldsymbol P(i+1)=\lambda^{-1}\boldsymbol P(i)-\lambda^{-1}\boldsymbol k(i)\boldsymbol x^H(i)\boldsymbol P(i).
\end{equation}
Following the adaptation step, we introduce the combination step for
both ES--LMS and ES--RLS algorithms, based on an exhaustive search
strategy. At first, we introduce a tentative set $\Omega_k$ using a
combinatorial approach described by
\begin{equation}
{{\Omega_k}\in 2^{\mathcal{N}_k}\backslash\varnothing,} \label{combine operate}
\end{equation}
{ where the set $\Omega_k$ is a nonempty set with $2^{\mathcal{N}_k}$
elements.} After the tentative set $\Omega_k$ is defined, we write the cost
function (\ref{cost_function}) for each node as
\begin{equation}
{J_{\boldsymbol\psi(i)}\big({\boldsymbol \psi(i)}\big)} \triangleq {\mathbb{E} \big|{ d_k(i)}-{\boldsymbol \psi}^H(i){\boldsymbol x_k(i)}\big|^2} ,
\end{equation}
where
\begin{equation}
{\boldsymbol \psi}(i) \triangleq \sum\limits_{l\in \Omega_k} c_{kl}(i) \boldsymbol\psi_l(i)
\end{equation}
is the local estimator and $\boldsymbol\psi_l(i)$ is calculated
through (\ref{LMS_adapt}) or (\ref{ES_RLS1}), depending on the
algorithm, i.e., ES--LMS or ES--RLS. With different
choices of the set $\Omega_k$, the combining coefficients $c_{kl}$
will be re--calculated through (\ref{Metropolis}), to ensure
condition (\ref{condition_sum1}) is satisfied.

Then, we introduce the error pattern for each node, which is defined as
\begin{equation}
{e_{\Omega_k}(i)} \triangleq { d_k(i)}-{\bigg[\sum\limits_{l\in
\Omega_k} c_{kl}(i)  \boldsymbol\psi_l(i)\bigg]}^H{\boldsymbol
x_k(i)}. \label{error pattern}
\end{equation}
For each node $k$, the strategy that finds the best set
$\Omega_k(i)$ must solve the following optimization problem:
\begin{equation}
{ \hat{\Omega}_k(i) = \arg\min_{\Omega_k \in 2^{N_k}\setminus\varnothing} |
e_{\Omega_k}(i)|.} \label{find_set}
\end{equation}
After all steps have been completed, the combination step in
(\ref{comb_strat}) is performed as described by
\begin{equation}
{\boldsymbol {\omega}}_k(i)= \sum\limits_{l\in \widehat{\Omega}_k(i)} c_{kl}(i) \boldsymbol\psi_l(i). \label{ES_LMS_adapt}
\end{equation}
At this stage, the main steps of the ES--LMS and ES--RLS algorithms
have been completed. The proposed ES--LMS and ES--RLS algorithms
find the set $\widehat{\Omega}_k(i)$ that minimizes the error
pattern in (\ref{error pattern}) and (\ref{find_set}) and then use
this set of nodes to obtain
${\boldsymbol{\omega}}_k(i)$ through (\ref{ES_LMS_adapt}). %At this point, the
%dynamic topology adaptation has been achieved, as at different
%iterations $i$, each node will have different combination results.
The ES--LMS/ES--RLS algorithms are briefly summarized as follows:
\begin{description}
\item[Step 1]Each node performs the adaptation through its local information based on the LMS or RLS algorithm.
\item[Step 2]Each node finds the best set $\Omega_k(i)$, which satisfies (\ref{find_set}).
\item[Step 3]Each node combines the information obtained from its best set of neighbors through (\ref{ES_LMS_adapt}).
\end{description}
The details of the proposed ES--LMS and ES--RLS
algorithms are shown in Tables \ref{table1} and \ref{table11}. When
the ES--LMS and ES--RLS algorithms are implemented in networks with
a large number of small and low--power sensors, the computational
complexity cost may become high, as the algorithm in (16) requires
an exhaustive search and needs more computations to examine all the
possible sets $\Omega_k(i)$ at each time instant.

\begin{table}\scriptsize
\centering \caption{The ES-LMS Algorithm}
\begin{tabular}{l}\hline
Initialize: ${\boldsymbol {\omega}}_k(0)$=0\\
For each time instant $i$=1,2, . . . , I\\
\ \ \ \ For each node $k$=1,2, \ldots, N\\
\ \ \ \ \ \ \ \ \ \ ${\boldsymbol {\psi}}_k(i)= {\boldsymbol {\omega}}_k(i-1)+{\mu}_k {\boldsymbol x_k(i)}[{ d_k(i)}-
{\boldsymbol \omega}_k^H(i-1){\boldsymbol x_k(i)}]^*$\\
\ \ \ \ end\\
\ \ \ \ For each node $k$=1,2, \ldots, N\\
\ \ \ \ \ \ \ \ \ \ find all possible sets of ${\Omega_k}$\\
\ \ \ \ \ \ \ \ \ \ ${e_{\Omega_k}(i)} = { d_k(i)}-[{\sum\limits_{l\in \Omega_k} c_{kl}(i) \boldsymbol\psi_l(i)}]^H{\boldsymbol x_k(i)} $\\
\ \ \ \ \ \ \ \ \ \ $\widehat{\Omega}_k(i)=\arg\min\limits_{\Omega_k}{|e_{\Omega_k}(i)|}$\\
\ \ \ \ \ \ \ \ \ \ ${\boldsymbol {\omega}}_k(i)= \sum\limits_{l\in \widehat{\Omega}_k(i)} c_{kl}(i) \boldsymbol\psi_l(i)$\\
\ \ \ \ end\\
end\\
\hline
\end{tabular}
\vskip -15pt
\label{table1}
\end{table}
\vspace{-0.5em}

\begin{table}\scriptsize
\centering \caption{The ES-RLS Algorithm}
\begin{tabular}{l}\hline
Initialize: ${\boldsymbol {\omega}}_k(0)$=0 \\
\ \ \ \ \ \ \ \ \ \ \ \ $\boldsymbol phi^{-1}(0)=\delta^{-1}\boldsymbol I, \delta=$ small positive constant\\
For each time instant $i$=1,2, . . . , I\\
\ \ \ \ For each node $k$=1,2, \ldots, N\\
\ \ \ \ \ \ \ \ \ \ $\boldsymbol\phi^{-1}(i)= \lambda^{-1}  \boldsymbol\phi^{-1}(i-1)$ \\
\ \ \ \ \ \ \ \ \ \ \ \ \ \ \ \ \ \ \ \ \ \ \ \ $- \dfrac {\lambda ^{-2}\boldsymbol\phi^{-1}(i-1) \boldsymbol x_k(i) \boldsymbol x_k^H(i)\boldsymbol\phi^{-1}(i-1)}{1+\lambda^{-1} \boldsymbol x_k^H(i) \boldsymbol\phi^{-1}(i-1)\boldsymbol x_k(i)}$\\
\ \ \ \ \ \ \ \ \ \ $\boldsymbol P(i)=\boldsymbol\phi^{-1}(i)$\\
\ \ \ \ \ \ \ \ \ \ $\boldsymbol k(i)= \dfrac {\lambda ^{-1}\boldsymbol P(i) \boldsymbol x_k(i)}{1+\lambda^{-1} \boldsymbol x_k^H(i)\boldsymbol P(i)\boldsymbol x_k(i)}$\\
\ \ \ \ \ \ \ \ \ \ ${\boldsymbol {\psi}}_k(i)= {\boldsymbol {\omega}}_k(i-1)+ {\boldsymbol k(i)}[{ d_k(i)}-{\boldsymbol \omega}_k^H(i-1){\boldsymbol x_k(i)}]^*$\\
\ \ \ \ \ \ \ \ \ \ $\boldsymbol P(i+1)=\lambda^{-1}\boldsymbol P(i)-\lambda^{-1}\boldsymbol k(i)\boldsymbol x_k^H(i)\boldsymbol P(i)$\\
\ \ \ \ end\\
\ \ \ \ For each node $k$=1,2, \ldots, N\\
\ \ \ \ \ \ \ \ \ \ find all possible sets of ${\Omega_k}$\\
\ \ \ \ \ \ \ \ \ \ ${e_{\Omega_k}(i)} = { d_k(i)}-[{\sum\limits_{l\in \Omega_k} c_{kl}(i) \boldsymbol\psi_l(i)}]^H{\boldsymbol x_k(i)}$\\
\ \ \ \ \ \ \ \ \ \ $\widehat{\Omega}_k(i)=\arg\min\limits_{\Omega_k}{|e_{\Omega_k}(i)|}$\\
\ \ \ \ \ \ \ \ \ \ ${\boldsymbol {\omega}}_k(i)= \sum\limits_{l\in \widehat{\Omega}_k(i)} c_{kl}(i) \boldsymbol\psi_l(i)$\\
\ \ \ \ end\\
end\\
\hline
\end{tabular}
\vskip -15pt
\label{table11}
\end{table}

\subsection{Sparsity--Inspired LMS/RLS Link Selection }
\vspace{-0.3em}

\begin{figure}[!htb]
\begin{center}
\vspace{-0.2em}
\def\epsfsize#1#2{0.7\columnwidth}
\epsfbox{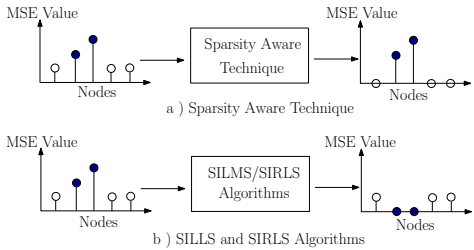} \vspace{-1.0em}\caption{\footnotesize Sparsity
aware signal processing strategies} \vskip -15pt \label{fig2}
\end{center}
\end{figure}

The ES--LMS/ES--RLS algorithms previously outlined need to examine all possible
sets to find a solution at each time instant, which might result in high
computational complexity for large networks operating in time--varying
scenarios. To solve the combinatorial problem with reduced complexity, we
propose sparsity-inspired based SI--LMS and SI--RLS algorithms, which are as
simple as standard diffusion LMS or RLS algorithms and are suitable for
adaptive implementations and scenarios where the parameters to be estimated are
slowly time--varying. The zero--attracting strategy (ZA), reweighted
zero--attracting strategy (RZA) and zero--forcing (ZF) are reported in
\cite{int,Chen,Meng,l1cg,zhaocheng,alt,jiolms,jiols,jiomimo,jidf,fa10,saabf,barc,honig,mswfccm,song,locsme}
as for sparsity aware techniques. These approaches are usually employed in
applications dealing with sparse systems in scenarios where they shrink the
small values in the parameter vector to zero, which results in better
convergence rate and steady--state performance. Unlike existing methods that
shrink the signal samples to zero, the proposed SI--LMS and SI--RLS algorithms
shrink to zero the links that have poor performance or high MSE values. To
detail the novelty of the proposed sparsity--inspired LMS/RLS link selection
algorithms, we illustrate the processing in Fig.\ref{fig2}.

Fig. \ref{fig2} (a) shows a standard type of sparsity--aware processing. We can
see that, after being processed by a sparsity--aware algorithm, the nodes with
small MSE values will be shrunk to zero. In contrast, the proposed SI--LMS and
SI--RLS algorithms will keep the nodes with lower MSE values and shrink the
nodes with large MSE values to zero as illustrated in Fig. \ref{fig2} (b). { In
the following, we will show how the proposed SI--LMS/SI--RLS algorithms are
employed to realize the link selection strategy automatically.}

In the adaptation step, we follow the same procedure in
(\ref{LMS_adapt})--(\ref{ES_RLS1}) as that of the ES--LMS and
ES--RLS algorithms for the SI--LMS and SI--RLS algorithms,
respectively. Then we reformulate the combination step.
First, we introduce the log--sum penalty into the
combination step in (\ref{comb_strat}). Different penalty terms have
been considered for this task. We have adopted a heuristic approach
\cite{Chen,Chen1} known as reweighted zero--attracting strategy into
the combination step in (\ref{comb_strat}) because this strategy has
shown an excellent performance and is simple to implement. The
regularization function with the log--sum penalty is defined as:
\begin{equation}
{f_1(e_{k}(i))}= \sum\limits_{l\in \mathcal{N}_k} \log\big(1+\varepsilon|e_{kl}(i)|\big),
\end{equation}
where the error pattern $e_{kl}(i) (l\in \mathcal{N}_k)$, which
stands for the neighbor node $l$ of node $k$ including node $k$
itself, is defined as
\begin{equation}
{e_{kl}(i)} \triangleq { d_k(i)}-{ \boldsymbol\psi_l^H(i)}{\boldsymbol x_k(i)} \label{error pattern2}
\end{equation}
and $\varepsilon$ is the shrinkage magnitude. Then, we
introduce the vector and matrix quantities required to describe the
combination step. We first define a vector $\boldsymbol c_k$ that
contains the combining coefficients for each neighbor of node $k$
including node $k$ itself as described by
\begin{equation}
{\boldsymbol c_k}\triangleq[c_{kl}],\ \ \ \ {l\in \mathcal{N}_k}.
\end{equation}
Then, we define a matrix $\boldsymbol \Psi_k$ that includes all the
estimated vectors, which are generated after the adaptation step of
SI--LMS and of SI--RLS for each neighbor of node $k$ including node
$k$ itself as given by
\begin{equation}
{\boldsymbol \Psi_k}\triangleq[\boldsymbol \psi_l(i)], \ \ \ \ {l\in
\mathcal{N}_k}.
\end{equation}
Note that the adaptation steps of SI--LMS and
SI--RLS are identical to (\ref{LMS_adapt}) and (\ref{ES_RLS1}),
respectively. An error vector $\hat{\boldsymbol e}_k$ that contains
all error values calculated through (\ref{error pattern2}) for each
neighbor of node $k$ including node $k$ itself is expressed by
\begin{equation}
{\hat{\boldsymbol e}_k}\triangleq[e_{kl}(i)], \ \ \ \ {l\in
\mathcal{N}_k}. \label{error vector}
\end{equation}
Here, we use a hat to distinguish $\hat{\boldsymbol
e}_k$ defined above from the original error $\boldsymbol e_k$. To
devise the sparsity--inspired approach, we have modified the vector
$\hat{\boldsymbol e}_k$ in the following way:
\begin{enumerate}
\item The element with largest absolute value $|e_{kl}(i)|$ in $\hat{\boldsymbol e}_k$ will be employed as $|e_{kl}(i)|$.
\item The element with smallest absolute value will be set to $-|e_{kl}(i)|$.
This process will ensure the node with smallest error pattern has a
reward on its combining coefficient.
\item The remaining entries will be set to zero.
\end{enumerate}
At this point, the combination step can be defined as \cite{Chen1}
\begin{equation}
\boldsymbol \omega_k(i)={\sum\limits_{j=1}^{|\mathcal{N}_k|}
\bigg[\boldsymbol c_k[j]-\rho \frac {\partial f_1(\hat{\boldsymbol
e}_k[j])}{\partial \hat{\boldsymbol e}_k[j]}\bigg]
\boldsymbol\Psi_k[j]}, \label{combination step3}
\end{equation}
where $\boldsymbol c_k[j], \hat{\boldsymbol e}_k[j]$ and
$\boldsymbol\Psi_k[j]$ stand for the $j$th element in the
$\boldsymbol c_k, \hat{\boldsymbol e}_k$ and $\boldsymbol\Psi_k$.
The parameter $\rho$ is used to control the algorithm's shrinkage
intensity. We then calculate the partial derivative
of $\hat{\boldsymbol e}_k[j]$:
\begin{align}
\frac{\partial{f_1(\hat{\boldsymbol e}_k[j])}}{\partial \hat{\boldsymbol e}_k[j]}&= \frac{\partial\big(\log(1+\varepsilon|e_{kl}(i)|)\big)}{\partial\big(e_{kl}(i)\big)}\notag\\
&=\varepsilon\frac{{\rm{sign}}(e_{kl}(i)) }{1+\varepsilon|e_{kl}(i)|}\ \ \ \ {l\in\mathcal{N}_k}\notag\\
&=\varepsilon\frac{{\rm{sign}}(\hat{\boldsymbol e}_k[j]) }{1+\varepsilon|\hat{\boldsymbol e}_k[j]|}. \label{pd}
\end{align}
To ensure that $\sum\limits_{j=1}^{|\mathcal{N}_k|}
\bigg(\boldsymbol c_k[j]-\rho \frac {\partial f_1(\hat{\boldsymbol
e}_k[j])}{\partial \hat{\boldsymbol e}_k[j]}\bigg)=1$, we replace
$\hat{\boldsymbol e}_k[j]$ with $\xi_{\rm min}$ in the denominator,
where the parameter $\xi_{\rm min}$ stands for the minimum absolute
value of $e_{kl}(i)$ in $\hat{\boldsymbol e}_k$. Then, (\ref{pd})
can be rewritten as
\begin{equation}
\frac{\partial{f_1(\hat{\boldsymbol e}_k[j])}}{\partial \hat{\boldsymbol e}_k[j]}=\varepsilon\frac{{\rm{sign}}(\hat{\boldsymbol e}_k[j]) }{1+\varepsilon|\xi_{\rm min}|} \label{partial derivative}.
\end{equation}
At this stage, the MSE cost function governs the
adaptation step, while the combination step employs the combining
coefficients with the derivative of the log-sum penalty which
performs shrinkage and selects the set of estimates from the
neighbor nodes with the best performance. The function
${\rm{sign}}(a)$ is defined as
\begin{equation}
{{\rm{sign}}(a)}=
\left\{\begin{array}{ll}
{a/|a|}\ \ \ \ \ {a\neq 0}\\
0\ \ \ \ \ \ \ \ \ \ {a= 0}.
\end{array}
\right.
\end{equation}
Then, by inserting (\ref{partial derivative}) into (\ref{combination
step3}), the proposed combination step is given by
\begin{equation}
\boldsymbol \omega_k(i)={\sum\limits_{j=1}^{|\mathcal{N}_k|}
\bigg[\boldsymbol c_k[j]-\rho
\varepsilon\frac{{\rm{sign}}({\hat{\boldsymbol e}_k[j]})
}{1+\varepsilon|\xi_{\rm min}|}\bigg] \boldsymbol\Psi_k[j]}
\label{combination step2}.
\end{equation}
Note that the condition $\boldsymbol c_k[j]-\rho
\varepsilon\frac{{\rm{sign}}({\hat{\boldsymbol e}_k[j]})
}{1+\varepsilon|\xi_{\rm min}|}\geq 0$ is enforced in
(\ref{combination step2}). When $\boldsymbol c_k[j]-\rho
\varepsilon\frac{{\rm{sign}}({\hat{\boldsymbol e}_k[j]})
}{1+\varepsilon|\xi_{\rm min}|}= 0$, it means that the corresponding
node has been discarded from the combination step. In the following
time instant, if this node still has the largest MSE, there will be
no changes in the combining coefficients for this set of nodes.

To guarantee the stability, the parameter $\rho$ is
assumed to be sufficiently small and the  penalty takes effect only
on the element in ${\hat{\boldsymbol e}_k}$ for which the magnitude
is comparable to $1/\varepsilon$ \cite{Chen}. Moreover, there is
little shrinkage exerted on the element in ${\hat{\boldsymbol e}_k}$
whose $|\hat{\boldsymbol e}_k[j]|\ll1/\varepsilon$. The SI--LMS and
SI--RLS algorithms perform link selection by the adjustment of the
combining coefficients through (\ref{combination step2}). At this
point, it should be emphasized that:
\begin{itemize}
\item The process in (\ref{combination step2}) satisfies condition (\ref{condition_sum1}), as the
penalty and reward amounts are the same for the nodes with maximum
and minimum error pattern, respectively.

\item When computing (\ref{combination step2}), there are no
matrix--vector multiplications. Therefore, no additional complexity
is introduced. As described in (\ref{combination step3}), only the
$j$th element in $\boldsymbol c_k, \hat{\boldsymbol e}_k$ and
$\boldsymbol\Psi_k$ are used for calculation.
\end{itemize}
For the neighbor node with the largest MSE value, after the
modifications of $\hat{\boldsymbol e}_k$, its $e_{kl}(i)$ value in
$\boldsymbol e_k$ will be a positive number which will lead to the
term $\rho\varepsilon\frac{{\rm{sign}}({\hat{\boldsymbol
e}_k[j]})}{1+\varepsilon|\xi_{\rm min}|}$ in (\ref{combination
step2}) being positive too. This means that the combining
coefficient for this node will be shrunk and the weight for this
node to build ${\boldsymbol {\omega}}_k(i)$ will be shrunk too. In
other words, when a node encounters high noise or interference
levels, the corresponding MSE value might be large. As a result, we
need to reduce the contribution of this group of nodes.

In contrast, for the neighbor node with the smallest MSE, as its
$e_{kl}(i)$ value in $\boldsymbol e_k$ will be a negative number,
the term $\rho\varepsilon\frac{{\rm{sign}}({\hat{\boldsymbol
e}_k[j]})}{1+\varepsilon|\xi_{\rm min}|}$ in (\ref{combination
step2}) will be negative too. As a result, the weight for this node
associated with the smallest MSE to build ${\boldsymbol
{\omega}}_k(i)$ will be increased. For the remaining neighbor nodes,
the entry $e_{kl}(i)$ in $\boldsymbol e_k$ is zero, which means the
term $\rho\varepsilon\frac{{\rm{sign}}({\hat{\boldsymbol
e}_k[j]})}{1+\varepsilon|\xi_{\rm min}|}$ in (\ref{combination
step2}) is zero and there is no change for the weights to build
${\boldsymbol {\omega}}_k(i)$. The main steps for the proposed
SI--LMS and SI--RLS algorithms are listed as follows:
\begin{description}
\item[Step 1]Each node carries out the adaptation through its local information based on the LMS or RLS algorithm.
\item[Step 2]Each node calculates the error pattern through (\ref{error pattern2}).
\item[Step 3]Each node modifies the error vector $\boldsymbol e_k$.
\item[Step 4]Each node combines the information obtained from its selected neighbors through (\ref{combination step2}).
\end{description}

The SI--LMS and SI--RLS algorithms are detailed in Table
\ref{table2}. For the ES--LMS/ES--RLS and SI--LMS/SI--RLS
algorithms, we design different combination steps and employ the
same adaptation procedure, which means the proposed algorithms have
the ability to equip any diffusion--type wireless networks operating
with other than the LMS and RLS algorithms. This includes, for example, the
diffusion conjugate gradient strategy \cite{Xu}.
\begin{table}\scriptsize
\centering \caption{The SI-LMS and SI-RLS Algorithms}
\begin{tabular}{l}
\hline
Initialize: ${\boldsymbol {\omega}}_k(-1)$=0 \\
\ \ \ \ \ \ \ \ \ \ \ \ $\boldsymbol P(0)=\delta^{-1}\boldsymbol I, \delta=$ small positive constant\\
For each time instant $i$=1,2, . . . , I\\
\ \ \ \ For each node $k$=1,2, \ldots, N\\
\ \ \ \ \ \ \ \ \ \ The adaptation step for computing $\boldsymbol\psi_k(i)$ \\
\ \ \ \ \ \ \ \ \ \ is exactly the same as the ES-LMS and ES-RLS\\
\ \ \ \ \ \ \ \ \ \ for the SI-LMS and SI-RLS algorithms respectively\\
\ \ \ \ end\\
\ \ \ \ For each node $k$=1,2, \ldots, N\\
\ \ \ \ \ \ \ \ \ \ ${e_{kl}(i)} = { d_k(i)}-{\boldsymbol x_k^H(i)}{ \boldsymbol\psi_l(i)}\ \ \ \ l\in \mathcal{N}_k$\\
\ \ \ \ \ \ \ \ \ \ ${\boldsymbol c_k}=[c_{kl}]\ \ \ \ {l\in \mathcal{N}_k}$\\
\ \ \ \ \ \ \ \ \ \ ${\boldsymbol \Psi_k}=[\boldsymbol \psi_l(i)]\ \ \ \ {l\in \mathcal{N}_k}$\\
\ \ \ \ \ \ \ \ \ \ ${\boldsymbol e_k}=[e_{kl}(i)]\ \ \ \ {l\in \mathcal{N}_k}$\\
\ \ \ \ \ \ \ \ \ \ Find the maximum and minimum absolute terms in ${\boldsymbol e_k}$\\
\ \ \ \ \ \ \ \ \ \ Modified ${\boldsymbol e_k}$ as ${\boldsymbol e_k}$=[0$\cdot\cdot\cdot$0,$\underbrace{|e_{kl}(i)|}_{\max}$,0$\cdot\cdot\cdot$0,$\underbrace{-|e_{kl}(i)|}_{\min}$,0$\cdot\cdot\cdot$0]\\
\ \ \ \ \ \ \ \ \ \ $\xi_{\rm min}= \min\big(|e_{kl}(i)|\big)$\\
\ \ \ \ \ \ \ \ \ \ $\boldsymbol \omega_k(i)={\sum\limits_{j=1}^{|\mathcal{N}_k|} \bigg[\boldsymbol c_k[j]-\rho
\varepsilon\frac{{\rm{sign}}({\boldsymbol e_k[j]})}{1+\varepsilon|\xi_{\rm min}|}\bigg] \boldsymbol\Psi_k[j]}$\\
\ \ \ \ end\\
end\\
\hline
\end{tabular}
\vskip -15pt
\label{table2}
\end{table}

\section{Analysis of the proposed algorithms}

In this section, a statistical analysis of the proposed algorithms
is developed, including a stability analysis and an MSE analysis of
the steady--state and tracking performance. In addition, the
computational complexity of the proposed algorithms is also
detailed. Before we start the analysis, we make some assumptions
that are common in the literature \cite{Haykin}.

\emph{Assumption I}: The weight-error vector
$\boldsymbol \varepsilon_k(i)$ and the input signal vector
$\boldsymbol x_k(i)$ are statistically independent, and the
weight--error vector for node $k$ is defined as
\begin{equation}
{\boldsymbol \varepsilon_k(i)} \triangleq \boldsymbol \omega_k(i) -
\boldsymbol \omega_0,
\end{equation}
where $\boldsymbol \omega_0$ denotes the optimum Wiener solution of
the actual parameter vector to be estimated, and $\boldsymbol
\omega_k(i)$ is the estimate produced by the proposed algorithms at
time instant $i$.

\emph{Assumption II}: The input signal vector $x_l(i)$ is drawn from
a stochastic process, which is ergodic in the autocorrelation
function \cite{Haykin}.

\emph{Assumption III}: The $M\times 1$ vector $\boldsymbol q(i)$
represents a stationary sequence of independent zero--mean vectors
and positive definite autocorrelation matrix $\boldsymbol
Q=\mathbb{E}[\boldsymbol q(i)\boldsymbol q^H(i)]$, which is independent of $\boldsymbol x_k(i)$, $\boldsymbol n_k(i)$ and
$\boldsymbol\varepsilon_l(i)$.

\emph{Assumption IV (Independence)}: All regressor input signals
$\boldsymbol x_k(i)$ are spatially and temporally independent.
\vspace{-0.5em}
\subsection{Stability Analysis}
\vspace{-0.3em} In general, to ensure that a
partially-connected network can converge to the global network
performance, information should be propagated across the network
\cite{Cattivelli33}. The work in \cite{Lopes11} shows that it is
central to the performance that each node should be able to reach
the other nodes through one or multiple hops \cite{Cattivelli33}. In
this section, we discuss the stability analysis of the proposed
ES--LMS and SI--LMS algorithms.

First, we substitute (\ref{LMS_adapt}) into
(\ref{ES_LMS_adapt}) and obtain
\begin{align}
{\boldsymbol {\omega}}_k(i+1)&= \sum\limits_{l\in \widehat{\Omega}_k(i)} c_{kl}(i) \boldsymbol\psi_l(i+1)\notag\\
&=\sum\limits_{l\in \widehat{\Omega}_k(i)} [{\boldsymbol {\omega}}_l(i)+{\mu}_l {\boldsymbol x_l(i+1)}e^*_l(i+1)]c_{kl}(i)\notag\\
&=\sum\limits_{l\in \widehat{\Omega}_k(i)} [\boldsymbol \omega_0+\boldsymbol\varepsilon_l(i)+{\mu}_l {\boldsymbol x_l(i+1)}e^*_l(i+1)]c_{kl}(i)\notag\\
&=\sum\limits_{l\in \widehat{\Omega}_k(i)} \boldsymbol \omega_0 c_{kl}+\sum\limits_{l\in \widehat{\Omega}_k(i)}[\boldsymbol\varepsilon_l(i)+{\mu}_l {\boldsymbol x_l(i+1)}e^*_l(i+1)]c_{kl}(i)\notag\\
& {\rm subject}\ {\rm to} \ \sum\limits_l c_{kl}(i)=1 \notag\\
&=\boldsymbol \omega_0+\sum\limits_{l\in \widehat{\Omega}_k(i)}[\boldsymbol\varepsilon_l(i)+{\mu}_l {\boldsymbol x_l(i+1)}e^*_l(i+1)]c_{kl}(i). \label{ES_LMS1}
\end{align}
Then, we have
\begin{align}
{\boldsymbol \varepsilon_k(i+1)}&= \boldsymbol \omega_0-\boldsymbol \omega_0+\sum\limits_{l\in \widehat{\Omega}_k(i)}[\boldsymbol\varepsilon_l(i)+{\mu}_l {\boldsymbol x_l(i+1)}e^*_l(i+1)]c_{kl}(i)\notag\\
&=\sum\limits_{l\in \widehat{\Omega}_k(i)}[\boldsymbol\varepsilon_l(i)+{\mu}_l{\boldsymbol x_l(i+1)}e^*_l(i+1)]c_{kl}(i).\label{ES_LMS222}
\end{align}
By employing \emph{Assumption IV}, we start with (\ref{ES_LMS222})
for the ES--LMS algorithm and define the global vectors and
matrices:
\begin{equation}
\boldsymbol\varepsilon(i+1) \triangleq [
\boldsymbol\varepsilon_1(i+1),\cdots,\boldsymbol\varepsilon_N(i+1)]^T
\end{equation}
\begin{equation}
\mathcal{\boldsymbol M} \triangleq \textrm{diag}\{\mu_1\boldsymbol
I_M,...,\mu_N\boldsymbol I_M\}
\end{equation}
\begin{equation}
\mathcal{\boldsymbol D}(i+1) \triangleq \textrm{diag}\{\boldsymbol
x_1(i+1)\boldsymbol x_1^H(i+1),...,\boldsymbol x_N(i+1)\boldsymbol
x_N^H(i+1)\}
\end{equation}
and the $NM \times 1$ vector
\begin{equation}
\mathcal{\boldsymbol g}(i+1) = [\boldsymbol x_1(i+1)n_1(i+1),\cdots,\boldsymbol x_N(i+1)n_N(i+1)]^T.
\end{equation}
We also define an $N\times N$ matrix $\boldsymbol C$ where the
combining coefficients \{$c_{kl}$\} correspond to the \{$l,k$\}
entries of the matrix $\boldsymbol C$ and the $NM \times NM$ matrix $\boldsymbol C_G$
with a Kronecker structure:
\begin{equation}
\boldsymbol C_G=\boldsymbol C\otimes \boldsymbol I_M
\end{equation}
where $\otimes$ denotes the Kronecker product.

By inserting $e_l(i+1)=e_{0-l}-\boldsymbol\varepsilon_l^H(i)\boldsymbol x_l(i+1)$
into (33), the global version of (\ref{ES_LMS222}) can then be
written as
\begin{equation}
\boldsymbol\varepsilon(i+1)=\boldsymbol C_G^T\big[\boldsymbol I
-\mathcal{\boldsymbol M} \mathcal{\boldsymbol D}(i+1)\big]
\boldsymbol\varepsilon(i)+\boldsymbol C_G^T\mathcal{\boldsymbol
M}\mathcal{\boldsymbol g}(i+1), \label{error_vec}
\end{equation}
where $e_{0-l}$ is the estimation error produced by the Wiener filter for node $l$ as
described by
\begin{equation}
e_{0-l}=d_l(i) - \boldsymbol \omega_0^H\boldsymbol x_l(i).\label{e_0}
\end{equation}
If we define
\begin{equation}
\begin{split}
\mathcal{D}& \triangleq \mathbb{E}[ \mathcal{\boldsymbol D}(i+1)]\\
&=\textrm{diag}\{\boldsymbol R_1,...,\boldsymbol R_N\}
\end{split}
\end{equation}
and take the expectation of (\ref{error_vec}), we arrive at
\begin{equation}
\mathbb{E}\{\boldsymbol\varepsilon(i+1)\} = \boldsymbol
C_G^T\big[\boldsymbol I-\mathcal{\boldsymbol
M}\mathcal{D}\big]\mathbb{E}\{\boldsymbol\varepsilon(i)\}.
\end{equation}
Before we proceed, let us define $\boldsymbol X=\boldsymbol
I-\mathcal{\boldsymbol M}\mathcal{D}$ and introduce \emph{Lemma 1}:

\emph{Lemma 1}: Let $\boldsymbol C_G$ and $\boldsymbol X$ denote
arbitrary $NM \times NM$ matrices, where $\boldsymbol C_G$ has real,
non-negative entries, with columns adding up to one. Then, the
matrix $\boldsymbol Y=\boldsymbol C_G^T\boldsymbol X$ is stable for
any choice of $\boldsymbol C_G$ if and only if $\boldsymbol X$ is
stable.

\emph{Proof}: Assume that  $\boldsymbol X$ is stable, it is true
that for every square matrix $\boldsymbol X$ and every $\alpha>0$,
there exists a submultiplicative matrix norm $||\cdot||_\tau$ that
satisfies $|| \boldsymbol X||_\tau\leq\tau(\boldsymbol X)+\alpha$,
where the submultiplicative matrix norm $||\boldsymbol A \boldsymbol
B ||\leq||\boldsymbol A ||\cdot|| \boldsymbol B||$ holds and
$\tau(\boldsymbol X)$ is the spectral radius of $\boldsymbol X$
\cite{Sayed,Kailath}. Since $\boldsymbol X$ is stable, $\tau(\boldsymbol X)<1$, and we can choose $\alpha>0$ such that $\tau(\boldsymbol X)+\alpha=v<1$
and $||\boldsymbol X||_\tau\leq v<1$. Then we obtain \cite{Cattivelli3}
\begin{equation}
\begin{split}
||\boldsymbol Y^i||_\tau&=||(\boldsymbol C_G^T\boldsymbol X)^i||_\tau\\
&\leq||(\boldsymbol C_G^T)^i||_{\tau}\cdot||\boldsymbol X^i||_\tau\\
&\leq v^i||(\boldsymbol C_G^T)^i||_\tau.
\end{split}
\end{equation}
Since $\boldsymbol C_G^T$ has non--negative entries with columns
that add up to one, it is element--wise bounded by unity. This means
its Frobenius norm is bounded as well and by the equivalence of
norms, so is any norm, in particular $||(\boldsymbol
C_G^T)^i||_{\tau}$. As a result, we have
\begin{equation}
\lim\limits_{i\rightarrow\infty}||\boldsymbol Y^i||_\tau=\textbf{0},
\end{equation}
where $\boldsymbol Y^i$ converges to the zero matrix for large $i$.
This establishes the stability of $(\boldsymbol C_G^T\boldsymbol
X)^i$.

A square matrix $\boldsymbol X$ is stable if it satisfies
$\boldsymbol X^i\rightarrow0$ as $i\rightarrow\infty$. In view of
\emph{Lemma 1} and (82), we need the matrix $\boldsymbol
I-\mathcal{\boldsymbol M}\mathcal{D}$ to be stable. As a result, it
requires $\boldsymbol I-\mu_k\boldsymbol R_k$ to be stable for all
$k$, which holds if the following condition is satisfied:
%Then, we take the expectation on both sides and obtain
%\begin{equation}
%\begin{split}
%{\mathbb{E}[\boldsymbol \varepsilon_k(i+1)}]&=\mathbb{E}\bigg[\sum\limits_{l\in \widehat{\Omega}_k(i)}[\boldsymbol\varepsilon_l(i)-{\mu}_l {\boldsymbol x_l(i+1)}e^*_l(i+1)]c_{kl}\bigg]\\
%&=\sum\limits_{l\in \widehat{\Omega}_k(i)}c_{kl}\mathbb{E}\bigg[\boldsymbol\varepsilon_l(i)-{\mu}_l {\boldsymbol x_l(i+1)}\bigg(d_l(i+1)\\
%&\ \ \ \ \ \ \ \ \ \ \ \ \ \ \ \ \ -\hat{\boldsymbol \omega}_l(i)^H\boldsymbol x_l(i+1)\bigg)^*\bigg]\\
%&=\sum\limits_{l\in \widehat{\Omega}_k(i)}c_{kl}\mathbb{E}\bigg[\boldsymbol\varepsilon_l(i)-\mu_l \boldsymbol x_l(i+1)e_0^*\\
%&\ \ \ \ \ \ \ \ \ \ \ \ \ \ \ \ \ -\mu_l\boldsymbol x_l(i+1)\boldsymbol x_l^H(i+1)\boldsymbol\varepsilon_l(i)\bigg].
%\end{split}
%\end{equation}
%From the orthogonality principle, we have
%\begin{equation}
%\sum\limits_{l\in \widehat{\Omega}_k(i)}c_{kl}\mathbb{E}[\mu_l \boldsymbol x_l(i+1)e_0^*]=0.
%\end{equation}
%Thus, we can simplify the recursion for the expectation of the
%weight--error vector $\boldsymbol \varepsilon_k(i+1)$ as
%\begin{equation}
%\mathbb{E}[\boldsymbol \varepsilon_k(i+1)]=\sum\limits_{l\in \widehat{\Omega}_k(i)}c_{kl}[I_M-\mu_l\boldsymbol R_l(i+1)]\mathbb{E}[\boldsymbol\varepsilon_l(i)].
%\end{equation}
%The mean $\boldsymbol \varepsilon_k(i+1)$ converges to zero as $i$ approaches infinity,

%\begin{equation}
%-1<\boldsymbol I-\mu_k\boldsymbol R_k<1, \label{SA1}
%\end{equation}
\begin{equation}
0<\mu_k<\dfrac {2} {\lambda_{max}\big(\boldsymbol R_k\big)} \label{SA}
\end{equation}
where $\lambda_{max}\big(\boldsymbol R_k\big)$ is the largest
eigenvalue of the correlation matrix $\boldsymbol R_k$. The
difference between the ES--LMS and SI--LMS algorithms is the
strategy to calculate the matrix $\boldsymbol C$. \emph{Lemma 1}
indicates that for any choice of $\boldsymbol C$, only $\boldsymbol
X$ needs to be stable. As a result, SI--LMS has the same convergence
condition as in (\ref{SA}). Given the convergence conditions, the
proposed ES--LMS/ES--RLS and SI--LMS/SI--RLS algorithms will adapt
according to the network connectivity by choosing the group of nodes
with the best available performance to construct their estimates.
Comparing the results in (\ref{SA}) with the existing algorithms, it
can be seen that the proposed link selection techniques change the
set of combining coefficients, which are indicated in $\boldsymbol
C_G$, as the matrix $\boldsymbol C$ employs the chosen set
$\widehat{\Omega}_k(i)$.

\subsection{MSE Steady--State Analysis}

In this part of the analysis, we devise formulas to
predict the excess MSE (EMSE) of the proposed algorithms. The error
signal at node $k$ can be expressed as
\begin{equation}
\begin{split}
{e_k(i)}&=d_k(i)-\boldsymbol \omega_k^H(i)\boldsymbol x_k(i)\\
&=d_k(i)-[\boldsymbol \omega_0 -\boldsymbol \varepsilon_k(i)]^H\boldsymbol x_k(i)\\
&=d_k(i)-\boldsymbol \omega_0^H\boldsymbol x_k(i) +\boldsymbol \varepsilon_k^H(i)\boldsymbol x_k(i)\\
&=e_{0-k}+\boldsymbol \varepsilon_k^H(i)\boldsymbol x_k(i).
\end{split}
\end{equation}
With \emph{Assumption I}, the MSE expression can be derived as
\begin{align}
{\mathcal{J}_{mse-k}(i)}&=\mathbb{E}[|e_k(i)|^2]\notag\\
&=\mathbb{E}\bigg[\big(e_{0-k}+\boldsymbol \varepsilon_k^H(i)\boldsymbol x_k(i)\big)\big(e_0^*+\boldsymbol x_k^H(i)\boldsymbol \varepsilon_k(i)\big)\bigg]\notag\\
&={\mathcal{J}_{min-k}}+\mathbb{E}[\boldsymbol \varepsilon_k^H(i)\boldsymbol x_k(i)\boldsymbol x_k^H(i)\boldsymbol \varepsilon_k(i)]\notag\\
&={\mathcal{J}_{min-k}}+\textrm{\textrm{tr}}\{\mathbb{E}[\boldsymbol \varepsilon_k(i)\boldsymbol \varepsilon_k^H(i)\boldsymbol x_k(i)\boldsymbol x_k^H(i)]\}\notag\\
&={\mathcal{J}_{min-k}}+\textrm{\textrm{tr}}\{\mathbb{E}[\boldsymbol x_k(i)\boldsymbol x_k^H(i)]\mathbb{E}[\boldsymbol \varepsilon_k(i)\boldsymbol \varepsilon_k^H(i)]\}\notag\\
&={\mathcal{J}_{min-k}}+\textrm{tr}\{\boldsymbol R_k(i)\boldsymbol K_k(i)\},
\end{align}
where $\textrm{tr}(\cdot)$ denotes the trace of a matrix and
$\mathcal{J}_{min-k}$ is the minimum mean--square error (MMSE) for
node $k$ \cite{Haykin}:
\begin{equation}
\mathcal{J}_{min-k}=\sigma_{d,k}^2-\boldsymbol p_k^H(i)\boldsymbol
R_k^{-1}(i)\boldsymbol p_k(i) \label{MMSE1},
\end{equation}
$\boldsymbol R_k(i)=\mathbb{E}[\boldsymbol x_k(i)\boldsymbol
x_k^H(i)]$ is the correlation matrix of the inputs for node $k$,
$\boldsymbol p_k(i)= \mathbb{E}[\boldsymbol x_k(i)d_k^*(i)]$ is the
cross--correlation vector between the inputs and the measurement
$d_k(i)$, and $\boldsymbol K_k(i)=\mathbb{E}[\boldsymbol
\varepsilon_k(i)\boldsymbol \varepsilon_k^H(i)]$ is the
weight--error correlation matrix. From \cite{Haykin}, the EMSE is
defined as the difference between the mean--square error at time
instant $i$ and the minimum mean--square error. Then, we can write
\begin{equation}
\begin{split}
{\mathcal{J}_{ex-k}(i)}&={\mathcal{J}_{mse-k}(i)}-{\mathcal{J}_{min-k}}\\
&=\textrm{tr}\{\boldsymbol R_k(i)\boldsymbol K_k(i)\}. \label{MMSE}
\end{split}
\end{equation}
For the proposed adaptive link selection algorithms, we will derive
the EMSE formulas separately based on (\ref{MMSE}) and \emph{we
assume that the input signal is modeled as a stationary process}.

\subsubsection{ES--LMS}

To update the estimate $\boldsymbol\omega_k(i)$, we
employ
\begin{align}
{\boldsymbol {\omega}}_k(i+1)&= \sum\limits_{l\in \widehat{\Omega}_k(i)} c_{kl}(i) \boldsymbol\psi_l(i+1)\notag\\
&=\sum\limits_{l\in \widehat{\Omega}_k(i)} c_{kl}(i)[{\boldsymbol {\omega}}_l(i)+{\mu}_l {\boldsymbol x_l(i+1)}e^*_l(i+1)]\notag\\
&=\sum\limits_{l\in \widehat{\Omega}_k(i)} c_{kl}(i)[{\boldsymbol {\omega}}_l(i)+{\mu}_l {\boldsymbol x_l(i+1)}(d_l(i+1)\notag\\
&\ \ \ \ \ \ \ \ \ \ \ \ \ \ \ \ \ \ \ \ -\boldsymbol x_l^H(i+1)\boldsymbol{\omega}_l(i))]. \label{eslms1}
\end{align}
Then, subtracting $\boldsymbol \omega_0$ from both sides of
(\ref{eslms1}), we arrive at
\begin{align}
{\boldsymbol {\varepsilon}}_k(i+1)&= \sum\limits_{l\in \widehat{\Omega}_k(i)} c_{kl}(i)[{\boldsymbol {\omega}}_l(i)+{\mu}_l {\boldsymbol x_l(i+1)}(d_l(i+1)\notag\\
&\ \ \ \ \ \ \ \ \ \ \ \ \ \ -\boldsymbol x_l^H(i+1)\boldsymbol{\omega}_l(i))]-\sum\limits_{l\in \widehat{\Omega}_k(i)} c_{kl}(i)\boldsymbol \omega_0\notag\\
&=\sum\limits_{l\in \widehat{\Omega}_k(i)} c_{kl}(i)\bigg[{\boldsymbol {\varepsilon}}_l(i)+{\mu}_l {\boldsymbol x_l(i+1)}\big(d_l(i+1)\notag\\
&\ \ \ \ \ \ \ \ \ \ \ \ \ \ -\boldsymbol x_l^H(i+1)(\boldsymbol {\varepsilon}_l(i)+\boldsymbol \omega_0)\big)\bigg]\notag\\
&=\sum\limits_{l\in \widehat{\Omega}_k(i)} c_{kl}(i)\bigg[{\boldsymbol {\varepsilon}}_l(i)+{\mu}_l {\boldsymbol x_l(i+1)}\big(d_l(i+1)\notag\\
&\ \ \ \ \ \ \ \ \ \ \ \ \ \ -\boldsymbol x_l^H(i+1)\boldsymbol {\varepsilon}_l(i)-\boldsymbol x_l^H(i+1)\boldsymbol \omega_0\big)\bigg]\notag\\
&=\sum\limits_{l\in \widehat{\Omega}_k(i)} c_{kl}(i)\bigg[{\boldsymbol {\varepsilon}}_l(i)-{\mu}_l {\boldsymbol x_l(i+1)}\boldsymbol x_l^H(i+1)\boldsymbol {\varepsilon}_l(i)\notag\\
&\ \ \ \ \ \ \ \ \ \ \ \ \ \ +{\mu}_l {\boldsymbol x_l(i+1)}e_{0-l}^*(i+1)\bigg]\notag\\
&=\sum\limits_{l\in \widehat{\Omega}_k(i)} c_{kl}(i)\bigg[\big(\boldsymbol I-{\mu}_l {\boldsymbol x_l(i+1)}\boldsymbol x_l^H(i+1)\big){\boldsymbol {\varepsilon}}_l(i)\notag\\
&\ \ \ \ \ \ \ \ \ \ \ \ \ \ +{\mu}_l {\boldsymbol x_l(i+1)}e_{0-l}^*(i+1)\bigg]. \label{eslms2}
\end{align}
Let us introduce the random variables $\alpha_{kl}(i)$:
\begin{equation}
\alpha_{kl}(i)=\left\{\begin{array}{ll}
1,\ \ $if$\ l\in\widehat{\Omega}_k(i)\\
0, \ \ $otherwise$.
\end{array}
\right.
\end{equation}
{  At each time instant, each node will generate data associated with network
covariance matrices $\boldsymbol A_k$ with size $N \times N$ which reflect the
network topology, according to the exhaustive search strategy. In the network
covariance matrices $\boldsymbol A_k$, a value equal to 1 means nodes $k$ and
$l$ are linked and a value 0 means nodes $k$ and $l$ are not linked.

For example, suppose a network has 5 nodes. For node 3, there are
two neighbor nodes, namely, nodes 2 and 5. Through Eq. (\ref{combine
operate}), the possible configurations of set $\Omega_3$ are $\{3,
2\}, \{3, 5\}$ and $\{3, 2, 5\}$. Evaluating all the possible sets
for $\Omega_3$, the relevant covariance matrices $\boldsymbol A_3$
with size $5\times5$ at time instant $i$ are described in Table
\ref{ex1}.

\begin{table}{
\caption{Covariance matrices $\boldsymbol A_3$ for different sets of $\Omega_3$}
\centering
\subtable[$\{3, 2\}$]{
       \begin{tabular}{|c|c|c|c|c|c|}
        \hline
        &1&2&3&4&5\\
        \hline
        1&&&0&&\\
        \hline
        2&&&1&&\\
        \hline
        3&0&1&1&0&0\\
        \hline
        4&&&0&&\\
        \hline
        5&&&0&&\\
        \hline
       \end{tabular}
}
\qquad
\subtable[$\{3, 5\}$ ]{
       \begin{tabular}{|c|c|c|c|c|c|}
        \hline
        &1&2&3&4&5\\
        \hline
        1&&&0&&\\
        \hline
        2&&&0&&\\
        \hline
        3&0&0&1&0&1\\
        \hline
        4&&&0&&\\
        \hline
        5&&&1&&\\
        \hline
       \end{tabular}
}
\qquad
\subtable[$\{3, 2, 5\}$ ]{
       \begin{tabular}{|c|c|c|c|c|c|}
        \hline
        &1&2&3&4&5\\
        \hline
        1&&&0&&\\
        \hline
        2&&&1&&\\
        \hline
        3&0&1&1&0&1\\
        \hline
        4&&&0&&\\
        \hline
        5&&&1&&\\
        \hline
       \end{tabular}
}\label{ex1}}
\end{table}

Then, the coefficients $\alpha_{kl}$ are obtained according to the
covariance matrices $\boldsymbol A_k$. In this example, the three
sets of $\alpha_{kl}$ are respectively shown in Table \ref{ex2}.}

\begin{table}{
\caption{Coefficients $\alpha_{kl}$ for different sets of $\Omega_3$}
\centering
\subtable[$\{3, 2\}$]{
$\left\{\begin{array}{lllll}
\alpha_{31}=0\\
\alpha_{32}=1\\
\alpha_{33}=1\\
\alpha_{34}=0\\
\alpha_{35}=0
\end{array}
\right. $

}
\qquad
\subtable[$\{3, 5\}$ ]{
$\left\{\begin{array}{lllll}
\alpha_{31}=0\\
\alpha_{32}=0\\
\alpha_{33}=1\\
\alpha_{34}=0\\
\alpha_{35}=1
\end{array}
\right. $
}
\qquad
\subtable[$\{3, 2, 5\}$ ]{
$\left\{\begin{array}{lllll}
\alpha_{31}=0\\
\alpha_{32}=1\\
\alpha_{33}=1\\
\alpha_{34}=0\\
\alpha_{35}=1
\end{array}
\right. $
}\label{ex2}}
\end{table}

{ The parameters $c_{kl}$ will then be calculated through Eq.
(\ref{Metropolis}) for different choices of matrices $\boldsymbol A_k$ and
coefficients $\alpha_{kl}$. After $\alpha_{kl}$ and $c_{kl}$ are calculated,
the error pattern for each possible $\Omega_k$ will be calculated through
(\ref{error pattern}) and the set with the smallest error will be selected
according to (\ref{find_set}).}

With the newly defined $\alpha_{kl}$, (\ref{eslms2}) can be rewritten as
\begin{align}
{\boldsymbol {\varepsilon}}_k(i+1)&=\sum\limits_{l\in \mathcal{N}_k} \alpha_{kl}(i)c_{kl}(i)\bigg[\big(\boldsymbol I-{\mu}_l {\boldsymbol x_l(i+1)}\boldsymbol x_l^H(i+1)\big){\boldsymbol {\varepsilon}}_l(i)\notag\\
&\ \ \ \ \ \ \ \ \ \ \ \ \ \ +{\mu}_l {\boldsymbol x_l(i+1)}e_{0-l}^*(i+1)\bigg]. \label{eslms3}
\end{align}
Starting from (\ref{MMSE}), we then focus on $\boldsymbol K_k(i+1)$.
\begin{align}
\boldsymbol K_k(i+1)&=\mathbb{E}[\boldsymbol\varepsilon_k(i+1)\boldsymbol \varepsilon_k^H(i+1)].\label{eslms4}
\end{align}
In (\ref{eslms3}), the term $\alpha_{kl}(i)$ is determined through
the network topology for each subset, while the term $c_{kl}(i)$ is
calculated through the Metropolis rule. We assume that
$\alpha_{kl}(i)$ and $c_{kl}(i)$ are statistically independent from
the other terms in (\ref{eslms3}). Upon convergence, the parameters
$\alpha_{kl}(i)$ and $c_{kl}(i)$ do not vary because at steady state
the choice of the subset $\widehat{\Omega}_k(i)$ for each node $k$
will be fixed. Then, under these assumptions, substituting
(\ref{eslms3}) into (\ref{eslms4}) we arrive at:
\begin{align}
\boldsymbol K_k(i+1)&=\sum\limits_{l\in \mathcal{N}_k} \mathbb{E}\bigg[\alpha_{kl}^2(i)c_{kl}^2(i)\bigg]\bigg(\big(\boldsymbol I-\mu_l\boldsymbol R_l(i+1)\big)\boldsymbol K_l(i)\notag\\
&\times\big(\boldsymbol I-\mu_l\boldsymbol R_l(i+1)\big)+\mu_l^2e_{0-l}(i+1)e_{0-l}^*(i+1)\notag\\
&\times\boldsymbol R_l(i+1)\bigg)+\sum\limits_{\substack{{l,q}\in \mathcal{N}_k \\ l\neq q}}\mathbb{E}\bigg[\alpha_{kl}(i)\alpha_{kq}(i)c_{kl}(i)c_{kq}(i)\bigg]\notag\\
&\times\bigg(\big(\boldsymbol I-\mu_l\boldsymbol R_l(i+1)\big)\boldsymbol K_{l,q}(i)\big(\boldsymbol I-\mu_q\boldsymbol R_l(i+1)\big)\notag\\
&+\mu_l\mu_qe_{0-l}(i+1)e_{0-q}^*(i+1)\boldsymbol R_{l,q}(i+1)\bigg)\notag\\
&+\sum\limits_{\substack{{l,q}\in \mathcal{N}_k \\ l\neq q}}\mathbb{E}\bigg[\alpha_{kl}(i)\alpha_{kq}(i)c_{kl}(i)c_{kq}(i)\bigg]\notag\\
&\times\bigg(\big(\boldsymbol I-\mu_q\boldsymbol R_q(i+1)\big)\boldsymbol K_{l,q}^H(i)\big(\boldsymbol I-\mu_l\boldsymbol R_l(i+1)\big)\notag\\
&+\mu_l\mu_qe_{0-q}(i+1)e_{0-l}^*(i+1)\boldsymbol R_{l,q}^H(i+1)\bigg)
\end{align}
where $\boldsymbol R_{l,q}(i+1)=\mathbb{E}[\boldsymbol
x_l(i+1)\boldsymbol x_q^H(i+1)]$ and $\boldsymbol
K_{l,q}(i)=\mathbb{E}[\boldsymbol\varepsilon_l(i)\boldsymbol
\varepsilon_q^H(i)]$. To further simplify the analysis, we assume
that the samples of the input signal $\boldsymbol x_k(i)$ are
uncorrelated, i.e., $\boldsymbol R_k=\sigma_{x,k}^2\boldsymbol I$
with $\sigma_{x,k}^2$ being the variance. Using the diagonal
matrices $\boldsymbol
R_k=\boldsymbol\Lambda_k=\sigma_{x,k}^2\boldsymbol I$ and
$\boldsymbol
R_{l,q}=\boldsymbol\Lambda_{l,q}=\sigma_{x,l}\sigma_{x,q}\boldsymbol
I$ we can write
\begin{align}
\boldsymbol K_k(i+1)&=\sum\limits_{l\in \mathcal{N}_k} \mathbb{E}\bigg[\alpha_{kl}^2(i)c_{kl}^2(i)\bigg]\bigg(\big(\boldsymbol I-\mu_l\boldsymbol\Lambda_l\big)\boldsymbol K_l(i)\big(\boldsymbol I-\mu_l\boldsymbol\Lambda_l\big)\notag\\
&+\mu_l^2e_{0-l}(i+1)e_{0-l}^*(i+1)\boldsymbol\Lambda_l\bigg)\notag\\
&+\sum\limits_{\substack{{l,q}\in \mathcal{N}_k \\ l\neq q}}
\mathbb{E}\bigg[\alpha_{kl}(i)\alpha_{kq}(i)c_{kl}(i)c_{kq}(i)\bigg] \notag\\
& \times \bigg(\big(\boldsymbol I-\mu_l\boldsymbol\Lambda_l\big)\boldsymbol K_{l,q}(i)\notag\\
&\times\big(\boldsymbol I-\mu_q\boldsymbol\Lambda_q\big) +\mu_l\mu_qe_{0-l}(i+1)e_{0-q}^*(i+1)\boldsymbol\Lambda_{l,q}\bigg)\notag\\
&+\sum\limits_{\substack{{l,q}\in \mathcal{N}_k \\ l\neq q}}\mathbb{E}\bigg[\alpha_{kl}(i)\alpha_{kq}(i)c_{kl}(i)c_{kq}(i)\bigg]\bigg(\big(\boldsymbol I-\mu_q\boldsymbol\Lambda_q\big)\boldsymbol K_{l,q}^H(i)\notag\\
&\times\big(\boldsymbol I-\mu_l\boldsymbol\Lambda_l\big) +\mu_l\mu_qe_{0-q}(i+1)e_{0-l}^*(i+1)\boldsymbol\Lambda_{l,q}^H\bigg).\label{eslms5}
\end{align}
Due to the structure of the above equations, the approximations and
the quantities involved, we can decouple (\ref{eslms5}) into
\begin{align}
K_k^n(i+1)&=\sum\limits_{l\in \mathcal{N}_k} \mathbb{E}\bigg[\alpha_{kl}^2(i)c_{kl}^2(i)\bigg]\bigg(\big(1-\mu_l\lambda_l^n\big)K_l^n(i)\big(1-\mu_l\lambda_l^n\big)\notag\\
&+\mu_l^2e_{0-l}(i+1)e_{0-l}^*(i+1)\lambda_l^n\bigg)\notag\\
&+\sum\limits_{\substack{{l,q}\in \mathcal{N}_k \\ l\neq q}}\mathbb{E}\bigg[\alpha_{kl}(i)\alpha_{kq}(i)c_{kl}(i)c_{kq}(i)\bigg]\bigg(\big(1-\mu_l\lambda_l^n\big)K_{l,q}^n(i)\notag\\
&\times\big(1-\mu_q\lambda_q^n\big) +\mu_l\mu_qe_{0-l}(i+1)e_{0-q}^*(i+1)\lambda_{l,q}^n\bigg)\notag\\
&+\sum\limits_{\substack{{l,q}\in \mathcal{N}_k \\ l\neq q}}\mathbb{E}\bigg[\alpha_{kl}(i)\alpha_{kq}(i)c_{kl}(i)c_{kq}(i)\bigg]\bigg(\big(1-\mu_q\lambda_q^n\big)(K_{l,q}^n)^H(i)\notag\\
&\times\big(1-\mu_l\lambda_l^n\big) +\mu_l\mu_qe_{0-q}(i+1)e_{0-l}^*(i+1)\lambda_{l,q}^n\bigg), \label{eslms6}
\end{align}
where $K_k^n(i+1)$ is the $n$th element of the main diagonal of
$\boldsymbol K_k(i+1)$. With the assumption that $\alpha_{kl}(i)$
and $c_{kl}(i)$ are statistically independent from the other terms
in (\ref{eslms3}), we can rewrite (\ref{eslms6}) as
\begin{align}
K_k^n(i+1)&=\sum\limits_{l\in \mathcal{N}_k} \mathbb{E}\bigg[\alpha_{kl}^2(i)\bigg]\mathbb{E}\bigg[c_{kl}^2(i)\bigg]\bigg(\big(1-\mu_l\lambda_l^n\big)^2K_l^n(i)\notag\\
&+\mu_l^2e_{0-l}(i+1)e_{0-l}^*(i+1)\lambda_l^n\bigg)\notag\\
&+2\times\sum\limits_{\substack{{l,q}\in \mathcal{N}_k \\ l\neq q}}\mathbb{E}\bigg[\alpha_{kl}(i)\alpha_{kq}(i)\bigg]\mathbb{E}\bigg[c_{kl}(i)c_{kq}(i)\bigg]\bigg(\big(1-\mu_l\lambda_l^n\big)\notag\\
&\times\big(1-\mu_q\lambda_q^n\big)K_{l,q}^n(i) +\mu_l\mu_qe_{0-l}(i+1)e_{0-q}^*(i+1)\lambda_{l,q}^n\bigg).
\end{align}
By taking $i\rightarrow\infty$, we can obtain (\ref{eslms7}).
\begin{figure*}[htp!]
\begin{equation}
\scriptsize{K_k^n(\text{ES-LMS})=\dfrac{\sum\limits_{l\in \mathcal{N}_k}\alpha_{kl}^2c_{kl}^2\mu_l^2\mathcal{J}_{min-l}\lambda_l^n+2\sum\limits_{\substack{{l,q}\in \mathcal{N}_k \\ l\neq q}}\alpha_{kl}\alpha_{kq}c_{kl}c_{kq}\mu_l\mu_qe_{0-l}e_{0-q}^*\lambda_{l,q}^n}{1-\sum\limits_{l\in \mathcal{N}_k}\alpha_{kl}^2c_{kl}^2(1-\mu_l\lambda_l^n)^2-2\sum\limits_{\substack{{l,q}\in \mathcal{N}_k \\ l\neq q}}\alpha_{kl}\alpha_{kq}c_{kl}c_{kq}(1-\mu_l\lambda_l^n)(1-\mu_q\lambda_q^n)}.} \label{eslms7}\vspace{-2.0em}
\end{equation}
\end{figure*}
{ It should be noticed that with the assumption that upon convergence the
choice of covariance matrix $\boldsymbol A_k$ for node $k$ is fixed, which
means it is deterministic and does not vary. In the above example, we assume
the choice of $\boldsymbol A_3$ is fixed as show in Table \ref{ex3}.
\begin{table}
\centering { \caption{Covariance matrix $\boldsymbol A_3$ upon convergence}
\begin{tabular}{|c|c|c|c|c|c|}
        \hline
        &1&2&3&4&5\\
        \hline
        1&&&0&&\\
        \hline
        2&&&1&&\\
        \hline
        3&0&1&1&0&1\\
        \hline
        4&&&0&&\\
        \hline
        5&&&1&&\\
        \hline
\end{tabular}\label{ex3}}
\end{table}

Then the coefficients $\alpha_{kl}$ will also be fixed and given by
\begin{equation}
\left\{\begin{array}{lllll}
\alpha_{31}=0\\
\alpha_{32}=1\\
\alpha_{33}=1\\
\alpha_{34}=0\\
\alpha_{35}=1
\end{array}
\right. \notag
\end{equation}
as well as the parameters $c_{kl}$ that are computed using the
Metropolis combining rule. As a result, the coefficients
$\alpha_{kl}$ and the coefficients $c_{kl}$ are deterministic and
can be taken out from the expectation.}

The MSE is then given by
\begin{equation}
\mathcal{J}_{mse-k}=\mathcal{J}_{min-k}+M\sigma_{x,k}^2\sum _{n=1}^{M}K_k^n(\text{ES-LMS}).\label{eslms8}
\end{equation}

\subsubsection{SI--LMS}

For the SI--LMS algorithm, we do not need to consider
all possible combinations. This algorithm simply adjusts the
combining coefficients for each node with its neighbors in order to
select the neighbor nodes that yield the smallest MSE values. Thus,
we redefine the combining coefficients through (\ref{combination
step2})
\begin{equation}
c_{kl-new}=c_{kl}-\rho\varepsilon\frac{{\rm{sign}}(|e_{kl}|)}{1+\varepsilon|\xi_{\rm min}|}\ \ (l\in\mathcal{N}_k). \label{new_c}
\end{equation}
{ For each node $k$, at time instant $i$, after it received the estimates from
all its neighbors, it calculates the error pattern $e_{kl}(i)$ for every
estimate received through Eq. (\ref{error pattern2}) and find the nodes with
the largest and smallest error. An error vector $ \hat{\boldsymbol e}_k$ is
then defined through (\ref{error vector}), which contains all error pattern
$e_{kl}(i)$ for node $k$.

Then a procedure which is detailed after Eq. (\ref{error vector}) is
carried out and modifies the error vector $ \hat{\boldsymbol e}_k$.
For example, suppose node $5$ has three neighbor nodes, which are
nodes $3, 6$ and $8$. The error vector $\hat{\boldsymbol e}_5$ has
the form described by $ \hat{\boldsymbol
e}_5=[e_{53},e_{55},e_{56},e_{58}]=[0.023,0.052,-0.0004,-0.012]$.
After the modification, the error vector $ \hat{\boldsymbol e}_5$
will be edited as $ \hat{\boldsymbol e}_5=[0,0.052,-0.0004,0]$. The
quantity $h_{kl}$ is then defined as
\begin{equation}
{ h_{kl}=\rho\varepsilon\frac{{\rm{sign}}(|e_{kl}|)}{1+\varepsilon|\xi_{\rm
min}|}\ \ (l\in\mathcal{N}_k), }\label{hh}
\end{equation}
and the term 'error pattern' $e_{kl}$ in (\ref{hh}) is from the
modified error vector $ \hat{\boldsymbol e}_k$.}

{  From \cite{Chen1}, we employ the relation
$\mathbb{E}[\textrm{sign}(e_{kl})]\approx \textrm{sign}(e_{0-k})$. According to
Eqs. (\ref{d_k}) and (\ref{e_0}), when the proposed algorithm converges at node
$k$ or the time instant $i$ goes to infinity, we assume that the error
$e_{0-k}$ will be equal to the noise variance at node $k$. Then, the asymptotic
value $h_{kl}$ can be divided into three situations due to the rule of the
SI--LMS algorithm:
\begin{equation}
{h_{kl}}=
\left\{\begin{array}{ll}
{\rho\varepsilon\frac{{\rm{sign}}(|e_{0-k}|)}{1+\varepsilon|e_{0-k}|}}\ \ \ \ $for the node with the largest MSE$\\
{\rho\varepsilon\frac{{\rm{sign}}(-|e_{0-k}|)}{1+\varepsilon|e_{0-k}|}}\ \ $for the node with the smallest MSE$\\
0\ \ \ \ \ \ \ \ \ \ \ \ \ \ \ $for all the remaining nodes$.
\label{h_{kl}}
\end{array}
\right.
\end{equation}
Under this situation, after the time instant $i$ goes to infinity,
the parameters $h_{kl}$ for each neighbor node of node $k$ can be
obtained through (\ref{h_{kl}}) and the quantity $h_{kl}$ will be
deterministic and can be taken out from the expectation.

At last, removing the random variables $\alpha_{kl}(i)$ and
inserting (\ref{new_c}), (\ref{hh}) into (\ref{eslms7}), the
asymptotic values $K_k^n$ for the SI-LMS algorithm are obtained as
in (\ref{silms1}).
\begin{figure*}[htp!]
\begin{equation}
\scriptsize{K_k^n(\text{SI-LMS})= \dfrac{\sum\limits_{l\in
\mathcal{N}_k}(c_{kl}-h_{kl})^2\mu_l^2\mathcal{J}_{min-l}\lambda_l^n+2\sum\limits_{\substack{{l,q}\in
\mathcal{N}_k \\ l\neq
q}}(c_{kl}-h_{kl})(c_{kq}-h_{kq})\mu_l\mu_qe_{0-l}e_{0-q}^*\lambda_{l,q}^n}{1-\sum\limits_{l\in
\mathcal{N}_k}(c_{kl}-h_{kl})^2(1-\mu_l\lambda_l^n)^2-2\sum\limits_{\substack{{l,q}\in
\mathcal{N}_k \\ l\neq
q}}(c_{kl}-h_{kl})(c_{kq}-h_{kq})(1-\mu_l\lambda_l^n)(1-\mu_q\lambda_q^n)}.}
\label{silms1}\vspace{-3.0em}
\end{equation}
\end{figure*}
At this point, the theoretical results are deterministic.}

Then, the MSE for SI--LMS algorithm is given by
\begin{equation}
\mathcal{J}_{mse-k}=\mathcal{J}_{min-k}+M\sigma_{x,k}^2\sum _{n=1}^{M}K_k^n(\text{SI-LMS}).\label{silms2}
\end{equation}

\subsubsection{ES--RLS}
For the proposed ES--RLS algorithm, we start from
(\ref{ES_RLS1}), after inserting (\ref{ES_RLS1}) into
(\ref{ES_LMS_adapt}), we have
\begin{align}
{\boldsymbol {\omega}}_k(i+1)&= \sum\limits_{l\in \widehat{\Omega}_k(i)} c_{kl}(i) \boldsymbol\psi_l(i+1)\notag\\
&=\sum\limits_{l\in \widehat{\Omega}_k(i)} c_{kl}(i)[{\boldsymbol {\omega}}_l(i)+\boldsymbol k_l(i+1)e^*_l(i+1)]\notag\\
&=\sum\limits_{l\in \widehat{\Omega}_k(i)} c_{kl}(i)[{\boldsymbol {\omega}}_l(i)+\boldsymbol k_l(i+1)(d_l(i+1)\notag\\
&\ \ \ \ \ \ \ \ \ \ \ \ \ \ \ \ \ \ \ \ -\boldsymbol x_l^H(i+1)\boldsymbol{\omega}_l(i))]. \label{esrls1}
\end{align}
Then, subtracting the $\boldsymbol \omega_0$ from both sides of (\ref{eslms1}), we arrive at
\begin{align}
{\boldsymbol {\varepsilon}}_k(i+1)&= \sum\limits_{l\in \widehat{\Omega}_k(i)} c_{kl}(i)[{\boldsymbol {\omega}}_l(i)+\boldsymbol k_l(i+1)(d_l(i+1)\notag\\
&\ \ \ \ \ \ \ \ \ \ \ \ \ \ -\boldsymbol x_l^H(i+1)\boldsymbol{\omega}_l(i))]-\sum\limits_{l\in \widehat{\Omega}_k(i)} c_{kl}(i)\boldsymbol \omega_0\notag\\
&=\sum\limits_{l\in \widehat{\Omega}_k(i)} c_{kl}(i)\bigg[{\boldsymbol {\varepsilon}}_l(i)+\boldsymbol k_l(i+1)\big(d_l(i+1)\notag\\
&\ \ \ \ \ \ \ \ \ \ \ \ \ \ -\boldsymbol x_l^H(i+1)(\boldsymbol {\varepsilon}_l(i)+\boldsymbol \omega_0)\big)\bigg]\notag\\
&=\sum\limits_{l\in \widehat{\Omega}_k(i)} c_{kl}(i)\bigg[\big(\boldsymbol I-\boldsymbol k_l(i+1)\boldsymbol x_l^H(i+1)\big){\boldsymbol {\varepsilon}}_l(i)\notag\\
&\ \ \ \ \ \ \ \ \ \ \ \ \ \ +\boldsymbol k_l(i+1)e_{0-l}^*(i+1)\bigg]. \label{esrls2}
\end{align}
Then, with the random variables $\alpha_{kl}(i)$, (\ref{esrls2}) can be rewritten as
\begin{align}
{\boldsymbol {\varepsilon}}_k(i+1)&=\sum\limits_{l\in \mathcal{N}_k} \alpha_{kl}(i)c_{kl}(i)\bigg[\big(\boldsymbol I-\boldsymbol k_l(i+1)\boldsymbol x_l^H(i+1)\big){\boldsymbol {\varepsilon}}_l(i)\notag\\
&\ \ \ \ \ \ \ \ \ \ \ \ \ \ +\boldsymbol k_l(i+1)e_{0-l}^*(i+1)\bigg]. \label{esrls3}
\end{align}
Since $\boldsymbol k_l(i+1)=\boldsymbol\Phi^{-1}_l(i+1)\boldsymbol
x_l(i+1)$ \cite{Haykin}, we can modify the (\ref{esrls3}) as
\begin{align}
{\boldsymbol {\varepsilon}}_k(i+1)&=\sum\limits_{l\in \mathcal{N}_k} \alpha_{kl}(i)c_{kl}(i)\bigg[\big(\boldsymbol I-\boldsymbol\Phi^{-1}_l(i+1)\boldsymbol x_l(i+1)\boldsymbol x_l^H(i+1)\big){\boldsymbol {\varepsilon}}_l(i)\notag\\
&\ \ \ \ \ \ \ \ \ \ \ \ \ \ +\boldsymbol\Phi^{-1}_l(i+1)\boldsymbol x_l(i+1)e_{0-l}^*(i+1)\bigg]. \label{esrls4}
\end{align}
At this point, if we compare (\ref{esrls4}) with (\ref{eslms3}), we
can find that  the difference between (\ref{esrls4}) and
(\ref{eslms3}) is, the $\boldsymbol\Phi^{-1}_l(i+1)$ in
(\ref{esrls4}) has replaced the ${\mu}_l$ in (\ref{eslms3}). From
\cite{Haykin}, we also have
\begin{equation}
\mathbb{E}[\boldsymbol \Phi^{-1}_l(i+1)]= \dfrac{1}{i-M}\boldsymbol R_l^{-1}(i+1)\ \ \ \ {\rm for}\ i>M+1.
\end{equation}
As a result, we can arrive
\begin{align}
\boldsymbol K_k(i+1)&=\sum\limits_{l\in \mathcal{N}_k} \mathbb{E}\bigg[\alpha_{kl}^2(i)c_{kl}^2(i)\bigg]\bigg(\big(\boldsymbol I-\dfrac{\boldsymbol\Lambda_l^{-1}\boldsymbol\Lambda_l}{i-M}\big)\boldsymbol K_l(i)\notag\\
&\times\big(\boldsymbol I-\dfrac{\boldsymbol\Lambda_l\boldsymbol\Lambda_l^{-1}}{i-M}\big)+\dfrac{\boldsymbol\Lambda_l^{-1}\boldsymbol\Lambda_l\boldsymbol\Lambda_l^{-1}}{(i-M)^2}e_{0-l}(i+1)e_{0-l}^*(i+1)\bigg)\notag\\
&+\sum\limits_{\substack{{l,q}\in \mathcal{N}_k \\ l\neq q}}\mathbb{E}\bigg[\alpha_{kl}(i)\alpha_{kq}(i)c_{kl}(i)c_{kq}(i)\bigg]\bigg(\big(\boldsymbol I-\dfrac{\boldsymbol\Lambda_l^{-1}\boldsymbol\Lambda_l}{i-M}\big)\notag\\
&\times\boldsymbol K_{l,q}(i)\big(\boldsymbol I-\dfrac{\boldsymbol\Lambda_q\boldsymbol\Lambda_q^{-1}}{i-M}\big) +\dfrac{\boldsymbol\Lambda_l^{-1}\boldsymbol\Lambda_{l,q}\boldsymbol\Lambda_q^{-1}}{(i-M)^2}e_{0-l}(i+1)\notag\\
&\times e_{0-q}^*(i+1)\bigg)+\sum\limits_{\substack{{l,q}\in \mathcal{N}_k \\ l\neq q}}\mathbb{E}\bigg[\alpha_{kl}(i)\alpha_{kq}(i)c_{kl}(i)c_{kq}(i)\bigg]\notag\\
&\times\bigg(\big(\boldsymbol I-\dfrac{\boldsymbol\Lambda_q\boldsymbol\Lambda_q^{-1}}{i-M}\big)\boldsymbol K_{l,q}^H(i)\big(\boldsymbol I-\dfrac{\boldsymbol\Lambda_l^{-1}\boldsymbol\Lambda_l}{i-M}\big) \notag\\
&+\dfrac{\boldsymbol\Lambda_q^{-1}\boldsymbol\Lambda_{l,q}^H\boldsymbol\Lambda_l^{-1}}{(i-M)^2}e_{0-q}(i+1)e_{0-l}^*(i+1)\bigg).\label{esrls5}
\end{align}
Due to the structure of the above equations, the approximations and
the quantities involved, we can decouple (\ref{esrls5}) into
\begin{align}
K_k^n(i+1)&=\sum\limits_{l\in \mathcal{N}_k} \mathbb{E}\bigg[\alpha_{kl}^2(i)c_{kl}^2(i)\bigg]\bigg(\big(1-\dfrac{1}{i-M}\big)^2 K_l^n(i)\notag\\
&+\dfrac{e_{0-l}(i+1)e_{0-l}^*(i+1)}{\lambda_l^{n}(i-M)^2}\bigg)\notag\\
&+\sum\limits_{\substack{{l,q}\in \mathcal{N}_k \\ l\neq q}}\mathbb{E}\bigg[\alpha_{kl}(i)\alpha_{kq}(i)c_{kl}(i)c_{kq}(i)\bigg]\bigg(\big(1-\dfrac{1}{i-M}\big)^2\notag\\
&\times K_{l,q}^n(i)+\dfrac{\lambda_{l,q}^n e_{0-l}(i+1)e_{0-q}^*(i+1)}{(i-M)^2\lambda_l^{n}\lambda_q^{n}}\bigg)\notag\\
&+\sum\limits_{\substack{{l,q}\in \mathcal{N}_k \\ l\neq q}}\mathbb{E}\bigg[\alpha_{kl}(i)\alpha_{kq}(i)c_{kl}(i)c_{kq}(i)\bigg]\bigg(\big(1-\dfrac{1}{i-M}\big)^2\notag\\
&\times (K_{l,q}^n(i))^H+\dfrac{\lambda_{l,q}^n e_{0-q}(i+1)e_{0-l}^*(i+1)}{(i-M)^2\lambda_q^{n}\lambda_l^{n}}\bigg)\label{esrls6}
\end{align}
where $K_k^n(i+1)$ is the $n$th elements of the main diagonals of
$\boldsymbol K_k(i+1)$. With the assumption that, upon convergence,
$\alpha_{kl}$ and $c_{kl}$ do not vary, because at steady state, the
choice of subset $\widehat{\Omega}_k(i)$ for each node $k$ will be
fixed, we can rewrite (\ref{esrls6}) as (\ref{esrls7}). Then, the
MSE is given by
\begin{figure*}[htp!]
\begin{equation}
\scriptsize{K_k^n(i+1)(\text{ES-RLS})=\dfrac{\sum\limits_{l\in
\mathcal{N}_k}\alpha_{kl}^2c_{kl}^2\dfrac{\mathcal{J}_{min-l}}{\lambda_l^n(i-M)^2}+2\sum\limits_{\substack{{l,q}\in
\mathcal{N}_k \\ l\neq
q}}\alpha_{kl}\alpha_{kq}c_{kl}c_{kq}\dfrac{\lambda_{l,q}^n
e_{0-l}e_{0-q}^*}{(i-M)^2\lambda_{l}^n\lambda_{q}^n}}{1-\sum\limits_{l\in
\mathcal{N}_k}\alpha_{kl}^2c_{kl}^2\Big(1-\dfrac{1}{i-M}\Big)^2-2\sum\limits_{\substack{{l,q}\in
\mathcal{N}_k \\ l\neq
q}}\alpha_{kl}\alpha_{kq}c_{kl}c_{kq}\Big(1-\dfrac{1}{i-M}\Big)^2}.}\label{esrls7}\vspace{-2.0em}
\end{equation}
\end{figure*}
\begin{equation}
\mathcal{J}_{mse-k}(i+1)=\mathcal{J}_{min-k}+M\sigma_{x,k}^2\sum _{n=1}^{M}K_k^n(i+1)(\text{ES-RLS}).\label{esrls8}
\end{equation}
On the basis of (\ref{esrls7}), we have that when $i$ tends to
infinity, the MSE approaches the MMSE in theory \cite{Haykin}.

\subsubsection{SI--RLS}
For the proposed SI--RLS algorithm, we insert (\ref{new_c}) into
(\ref{esrls7}), remove the random variables $\alpha_{kl}(i)$, and
following the same procedure as for the SI--LMS algorithm, we can
obtain (\ref{sirls1}), where $h_{kl}$ and $h_{kq}$ satisfy the rule
in (\ref{h_{kl}}). Then, the MSE is given by
\begin{figure*}[htp!]
\begin{equation}
\scriptsize{K_k^n(i+1)(\text{SI-RLS})=
\dfrac{\sum\limits_{l\in
\mathcal{N}_k}(c_{kl}-h_{kl})^2\dfrac{\mathcal{J}_{min-l}}{\lambda_l^n(i-M)^2}+2\sum\limits_{\substack{{l,q}\in
\mathcal{N}_k \\ l\neq
q}}(c_{kl}-h_{kl})(c_{kq}-h_{kq})\dfrac{\lambda_{l,q}^n
e_{0-l}e_{0-q}^*}{(i-M)^2\lambda_{l}^n\lambda_{q}^n}}{1-\sum\limits_{l\in
\mathcal{N}_k}(c_{kl}-h_{kl})^2\Big(1-\dfrac{1}{i-M}\Big)^2-2\sum\limits_{\substack{{l,q}\in
\mathcal{N}_k \\ l\neq
q}}(c_{kl}-h_{kl})(c_{kq}-h_{kq})\Big(1-\dfrac{1}{i-M}\Big)^2}.}
\label{sirls1}\vspace{-3.0em}
\end{equation}
\end{figure*}
\begin{equation}
\mathcal{J}_{mse-k}(i+1)=\mathcal{J}_{min-k}+M\sigma_{x,k}^2\sum _{n=1}^{M}K_k^n(i+1)(\text{SI-RLS}).\label{sirls2}
\end{equation}
In conclusion, according to (\ref{silms1}) and (\ref{sirls1}), with
the help of modified combining coefficients, for the proposed
SI--type algorithms, the neighbor node with lowest MSE contributes
the most to the combination, while the neighbor node with the
highest MSE contributes the least. Therefore, the proposed SI--type
algorithms perform better than the standard diffusion algorithms
with fixed combining coefficients.

\subsection{Tracking Analysis}

In this section, we assess the proposed ES--LMS/RLS and SI--LMS/RLS
algorithms in a non--stationary environment, in which the algorithms
have to track the minimum point of the error--performance surface
\cite{Rcdl3,Cai}. In the time--varying scenarios of interest, the
optimum estimate is assumed to vary according to the model
$\boldsymbol\omega_0(i+1)=\beta\boldsymbol\omega_0(i)+\boldsymbol
q(i)$, where $\boldsymbol q(i)$ denotes a random perturbation
\cite{Sayed} and $\beta=1$ in order to facilitate the analysis. This
is typical in the context of tracking analysis of adaptive
algorithms \cite{Haykin,Widrow,Eweda}.

\subsubsection{ES--LMS}

For the tracking analysis of the ES--LMS algorithm, we employ
\emph{Assumption III} and start from (\ref{eslms1}). After
subtracting the $\boldsymbol \omega_0(i+1)$ from both sides of
(\ref{eslms1}), we obtain
\begin{align}
{\boldsymbol {\varepsilon}}_k(i+1)&= \sum\limits_{l\in \widehat{\Omega}_k(i)} c_{kl}(i)[{\boldsymbol {\omega}}_l(i)+{\mu}_l {\boldsymbol x_l(i+1)}(d_l(i+1)\notag\\
&\ \ \ \ \ \ \ \ \ \ \ \ \ \ -\boldsymbol x_l^H(i+1)\boldsymbol{\omega}_l(i))]-\sum\limits_{l\in \widehat{\Omega}_k(i)} c_{kl}(i)\boldsymbol \omega_0(i+1)\notag\\
&= \sum\limits_{l\in \widehat{\Omega}_k(i)} c_{kl}(i)[{\boldsymbol {\omega}}_l(i)+{\mu}_l {\boldsymbol x_l(i+1)}(d_l(i+1)\notag\\
&\ \ \ \ \ \ \ \ \ \ -\boldsymbol x_l^H(i+1)\boldsymbol{\omega}_l(i))]-\sum\limits_{l\in \widehat{\Omega}_k(i)} c_{kl}(i)\bigg(\boldsymbol \omega_0(i)+\boldsymbol q(i)\bigg)\notag\\
&=\sum\limits_{l\in \widehat{\Omega}_k(i)} c_{kl}(i)\bigg[{\boldsymbol {\varepsilon}}_l(i)+{\mu}_l {\boldsymbol x_l(i+1)}\big(d_l(i+1)\notag\\
&\ \ \ \ \ \ \ \ \ \ \ \ \ \ -\boldsymbol x_l^H(i+1)(\boldsymbol {\varepsilon}_l(i)+\boldsymbol \omega_0)\big)\bigg]-\boldsymbol q(i)\notag\\
&=\sum\limits_{l\in \widehat{\Omega}_k(i)} c_{kl}(i)\bigg[\big(\boldsymbol I-{\mu}_l {\boldsymbol x_l(i+1)}\boldsymbol x_l^H(i+1)\big){\boldsymbol {\varepsilon}}_l(i)\notag\\
&\ \ \ \ \ \ \ \ \ \ \ \ \ \ +{\mu}_l {\boldsymbol x_l(i+1)}e_{0-l}^*(i+1)\bigg]-\boldsymbol q(i). \label{eslms2-2}
\end{align}
Using \emph{Assumption III}, we can arrive at
\begin{equation}
{\mathcal{J}_{ex-k}(i+1)}=\textrm{tr}\{\boldsymbol R_k(i+1)\boldsymbol K_k(i+1)\}+\textrm{tr}\{\boldsymbol R_k(i+1)\boldsymbol Q\}. \label{TA_ES_LMS}
\end{equation}
The first part on the right side of (\ref{TA_ES_LMS}), has already
been obtained in the MSE steady--state analysis part in Section IV
B. The second part can be decomposed as
\begin{align}
\textrm{tr}\{\boldsymbol R_k(i+1)\boldsymbol Q\}&=\textrm{tr}\bigg\{\mathbb{E}\big[\boldsymbol x_k(i+1)\boldsymbol x_k^H(i+1)\big]\mathbb{E}\big[\boldsymbol q(i)\boldsymbol q^H(i)\big]\bigg\}\notag\\
&=M\sigma_{x,k}^2\textrm{tr}\{\boldsymbol Q\}.
\end{align}
The MSE is then obtained as
\begin{equation}
\mathcal{J}_{mse-k}=\mathcal{J}_{min-k}+M\sigma_{x,k}^2\sum _{n=1}^{M}K_k^n(\text{ES-LMS})+M\sigma_{x,k}^2\textrm{tr}\{\boldsymbol Q\}.\label{eslms9}
\end{equation}

\subsubsection{SI--LMS}
For the SI--LMS recursions, we follow the same procedure as for the
ES-LMS algorithm, and obtain
\begin{equation}
\mathcal{J}_{mse-k}=\mathcal{J}_{min-k}+M\sigma_{x,k}^2\sum _{n=1}^{M}K_k^n(\text{SI-LMS})+M\sigma_{x,k}^2\textrm{tr}\{\boldsymbol Q\}.\label{silms3}
\end{equation}

\subsubsection{ES--RLS}
For the SI--RLS algorithm, we follow the same procedure as for the
ES--LMS algorithm and arrive at
\begin{align}
\mathcal{J}_{mse-k}(i+1)&=\mathcal{J}_{min-k}+M\sigma_{x,k}^2\sum _{n=1}^{M}K_k^n(i+1)(\text{ES-RLS})\notag\\
&\ \ \ +M\sigma_{x,k}^2\textrm{tr}\{\boldsymbol Q\}.\label{esrls9}
\end{align}

\subsubsection{SI--RLS}
We start from (\ref{sirls2}), and after a similar procedure to that of the SI--LMS algorithm, we have
\begin{align}
\mathcal{J}_{mse-k}(i+1)&=\mathcal{J}_{min-k}+M\sigma_{x,k}^2\sum _{n=1}^{M}K_k^n(i+1)(\text{SI-RLS})\notag\\
&\ \ \ +M\sigma_{x,k}^2\textrm{tr}\{\boldsymbol Q\}.\label{sirls3}
\end{align}
In conclusion, for time-varying scenarios there is only one
additional term $M\sigma_{x,k}^2\textrm{tr}\{\boldsymbol Q\}$ in the
MSE expression for all algorithms, and this additional term has the
same value for all algorithms. As a result, the proposed SI--type
algorithms still perform better than the standard diffusion
algorithms with fixed combining coefficients, according to the
conclusion obtained in the previous subsection.

\subsection{Computational Complexity}

In the analysis of the computational cost of the algorithms studied,
we assume complex-valued data and first analyze the adaptation step.
For both ES--LMS/RLS and SI--LMS/RLS algorithms, the adaptation cost
depends on the type of recursions (LMS or RLS) that each strategy
employs. The details are shown in Table \ref{table3}.
\begin{table}\scriptsize
\centering \caption{ \scriptsize{Computational complexity for the adaptation step
per node per time instant}}
\begin{tabular}{cccc}
\hline
Adaptation Method&Multiplications&Additions&Divisions\\
\hline
LMS&$8M+2$&$8M$&\\
RLS&$4M^2+16M+1$&$4M^2+12M-1$&$1$\\
\hline
\end{tabular}
\vskip -15pt \label{table3}
\end{table}
For the combination step, we analyze the
computational complexity in Table \ref{table4}. The overall
complexity for each algorithm is summarized in Table \ref{table5}.
In the above three tables, $T$ is the total number of nodes linked
to node $k$ including node $k$ itself and $t$ is the number of nodes
chosen from $T$. The proposed algorithms require extra computations
as compared to the existing distributed LMS and RLS algorithms. This
extra cost ranges from a small additional number of operations for
the SI-LMS/RLS algorithms to a more significant extra cost that
depends on $T$ for the ES-LMS/RLS algorithms.

\begin{table}\scriptsize
\centering \caption{\scriptsize{Computational complexity for combination step
per node per time instant}}
\begin{tabular}{cccc}
\hline
Algorithms&Multiplications&Additions&Divisions\\
\hline
ES--LMS/RLS&$M(t+1){\dfrac{T!}{t!(T-t)!}}$&$Mt{\dfrac{T!}{t!(T-t)!}}$&\\
SI--LMS/RLS&$(2M+3)|\mathcal{N}_k|$&$(M+2)|\mathcal{N}_k|$&$|\mathcal{N}_k|$\\
\hline
\end{tabular}
\vskip -10pt \label{table4}
\end{table}

\begin{table*}[!htb]\scriptsize
\centering \caption{\scriptsize{Computational complexity per node per time instant}}
\begin{tabular}{cccc}
\hline
Algorithm&Multiplications&Additions&Divisions\\
\hline
ES--LMS&$\bigg[{\dfrac{(t+1)T!}{t!(T-t)!}}+8\bigg]M+2$&$\bigg[{\dfrac{T!}{(t-1)!(T-t)!}}+8\bigg]M$&\\
ES--RLS&$4M^2+\bigg[{\dfrac{(t+1)T!}{t!(T-t)!}}+16\bigg]M+1$&$4M^2+\bigg[{\dfrac{T!}{(t-1)!(T-t)!}}+12\bigg]M-1$&$1$\\
SI--LMS&$(8+2\mathcal{N}_k)M+3|\mathcal{N}_k|+2$&$(8+|\mathcal{N}_k|)M+2|\mathcal{N}_k|$&$|\mathcal{N}_k|$\\
SI--RLS&$4M^2+(16+2|\mathcal{N}_k|)M+3|\mathcal{N}_k|+1$&$4M^2+(12+|\mathcal{N}_k|)M+2|\mathcal{N}_k|-1$&$|\mathcal{N}_k|+1$\\
\hline
\end{tabular}
\label{table5}\vspace{-2.5em}
\end{table*}

\section{Simulations}

{ In this section, we investigate the performance of the proposed link
selection strategies for distributed estimation in two scenarios: wireless
sensor networks and smart grids. In these applications, we simulate the
proposed link selection strategies in both static and time--varying scenarios.
We also show the analytical results for the MSE steady--state and tracking
performances that we obtained in Section IV.}

\subsection{Diffusion Wireless Sensor Networks}
\begin{figure}
\begin{center}
%\vspace{-2.0em}
\def\epsfsize#1#2{0.8\columnwidth}
\epsfbox{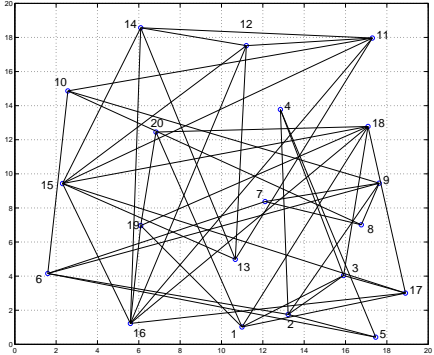}\vspace{-1.25em}\caption{\footnotesize
Diffusion wireless sensor networks topology with 20 nodes} \vskip
-20pt \label{fig3}
\end{center}
\end{figure}

In this subsection, we compare the proposed ES--LMS/ES--RLS and
SI--LMS/SI--RLS algorithms with the diffusion LMS algorithm
\cite{Lopes2}, the diffusion RLS algorithm \cite{Cattivelli2} and
the single--link strategy \cite{Zhao} in terms of their MSE
performance. The network topology is illustrated in Fig. \ref{fig3}
and we employ $N=20$ nodes in the simulations. The length of the
unknown parameter vector ${\boldsymbol \omega}_0$ is $M=10$ and it
is generated randomly. The input signal is generated as
${\boldsymbol x_k(i)}=[x_k(i)\ \ \ x_k(i-1)\ \ \ ...\ \ \
x_k(i-M+1)]$  and $x_k(i)=u_k(i)+\alpha_kx_k(i-1)$, where $\alpha_k$
is a correlation coefficient and $u_k(i)$ is a white noise process
with variance $\sigma^2_{u,k}= 1-|\alpha_k|^2$, to ensure the
variance of ${\boldsymbol x_k(i)}$ is $\sigma^2_{x,k}= 1$. The noise
samples are modeled as circular Gaussian noise with zero mean and
variance $\sigma^2_{n,k}= 0.001$. The step size for the diffusion
LMS ES--LMS and SI--LMS algorithms is $\mu=0.045$. For the diffusion
RLS algorithm, both ES--RLS and SI--RLS, the forgetting factor
$\lambda$ is set to 0.97 and $\delta$ is equal to 0.81. In the
static scenario, the sparsity parameters of the SI--LMS/SI--RLS
algorithms are set to $\rho=4\times10^{-3}$ and $\varepsilon=10$.
The Metropolis rule is used to calculate the combining coefficients
$c_{kl}$. The MSE and MMSE are defined as in (\ref{cost_function})
and (\ref{MMSE1}), respectively. The results are averaged over 100
independent runs.

\begin{figure}
\begin{center}
\def\epsfsize#1#2{0.8\columnwidth}
\epsfbox{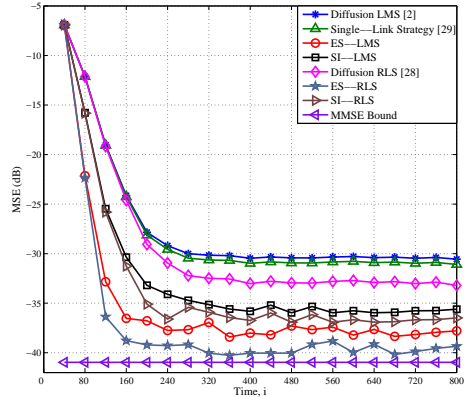}\vspace{-1.25em}\caption{\footnotesize Network
MSE curves in a static scenario} \vskip -20pt \label{fig4}
\end{center}
\end{figure}

In Fig. \ref{fig4}, we can see that ES--RLS has the best performance
for both steady-state MSE and convergence rate, and obtains a gain
of about 8 dB over the standard diffusion RLS algorithm. SI--RLS is
a bit worse than the ES--RLS, but is still significantly better than
the standard diffusion RLS algorithm by about 5 dB. Regarding the
complexity and processing time, SI--RLS is as simple as the standard
diffusion RLS algorithm, while ES--RLS is more complex. The proposed
ES--LMS and SI--LMS algorithms are superior to the standard
diffusion LMS algorithm. In the time--varying scenario, the sparsity
parameters of the SI--LMS and SI--RLS algorithms are set to
$\rho=6\times10^{-3}$ and $\varepsilon=10$. The unknown parameter
vector ${\boldsymbol {\omega}}_0$ varies according to the
first--order Markov vector process:
\begin{equation}
{\boldsymbol \omega}_0(i+1)=\beta{\boldsymbol
\omega}_0(i)+{\boldsymbol z(i)}, \label{tv}
\end{equation}
where $\boldsymbol z(i)$ is an independent zero--mean Gaussian
vector process with variance $\sigma^2_{z}= 0.01$ and $\beta=0.98$.

\begin{figure}
\begin{center}
\def\epsfsize#1#2{0.75\columnwidth}
\epsfbox{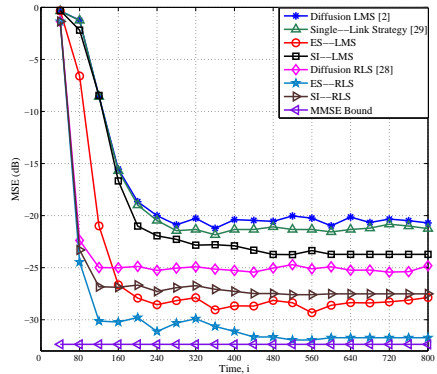}\vspace{-1.55em}\caption{\footnotesize
Network MSE curves in a time–-varying scenario}
\vskip -20pt
\label{fig5}
\end{center}
\end{figure}

\begin{figure}
\begin{center}
\def\epsfsize#1#2{0.85\columnwidth}
\epsfbox{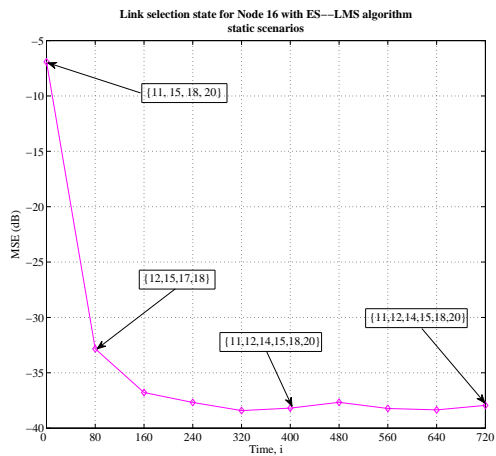}\vspace{-1.25em}\caption{\footnotesize Set of
selected links for node 16 with ES--LMS in a static scenario} \vskip
-20pt \label{link}
\end{center}
\end{figure}

\begin{figure}
\begin{center}
\def\epsfsize#1#2{0.875\columnwidth}
\epsfbox{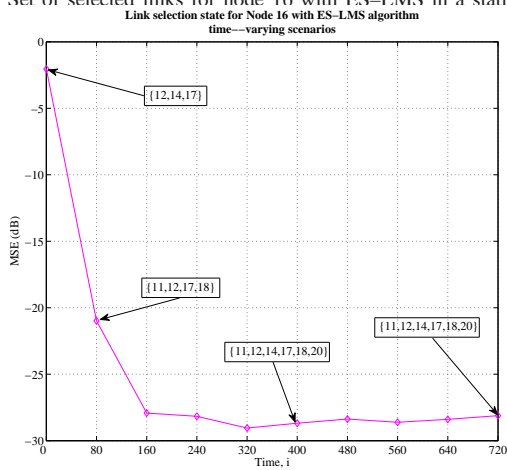}\vspace{-1.25em}\caption{\footnotesize Link
selection state for node 16 with ES--LMS in a time–-varying
scenario} \vskip -20pt \label{link1}
\end{center}
\end{figure}

Fig. \ref{fig5} shows that, similarly to the static scenario,
ES--RLS has the best performance, and obtains a 5 dB gain over the
standard diffusion RLS algorithm. SI--RLS is slightly worse than the
ES--RLS, but is still better than the standard diffusion RLS
algorithm by about 3 dB. The proposed ES--LMS and SI--LMS algorithms
have the same advantage over the standard diffusion LMS algorithm in
the time-varying scenario. Notice that in the scenario
with large $|\mathcal{N}_k|$, the proposed SI-type algorithms still
have a better performance when compared with the standard
techniques. To illustrate the link selection for the ES--type
algorithms, we provide Figs. \ref{link} and \ref{link1}. From these
two figures, we can see that upon convergence the proposed
algorithms converge to a fixed selected set of links
$\widehat{\Omega}_k$. \vspace{-0.5em}

\subsection{MSE Analytical Results}
\begin{figure}
\centerline{
\includegraphics[width=0.22\textwidth,height=0.45\textwidth]{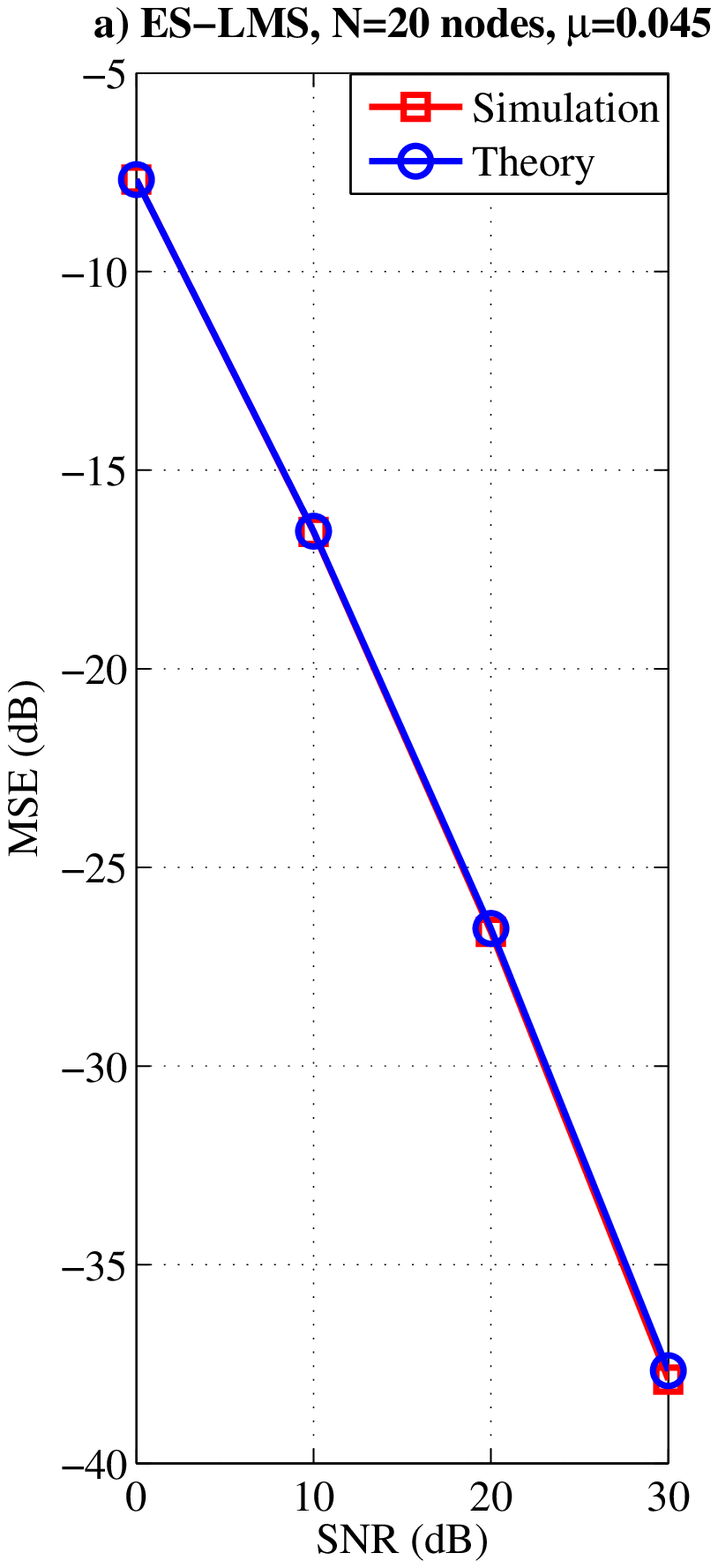}
\includegraphics[width=0.22\textwidth,height=0.45\textwidth]{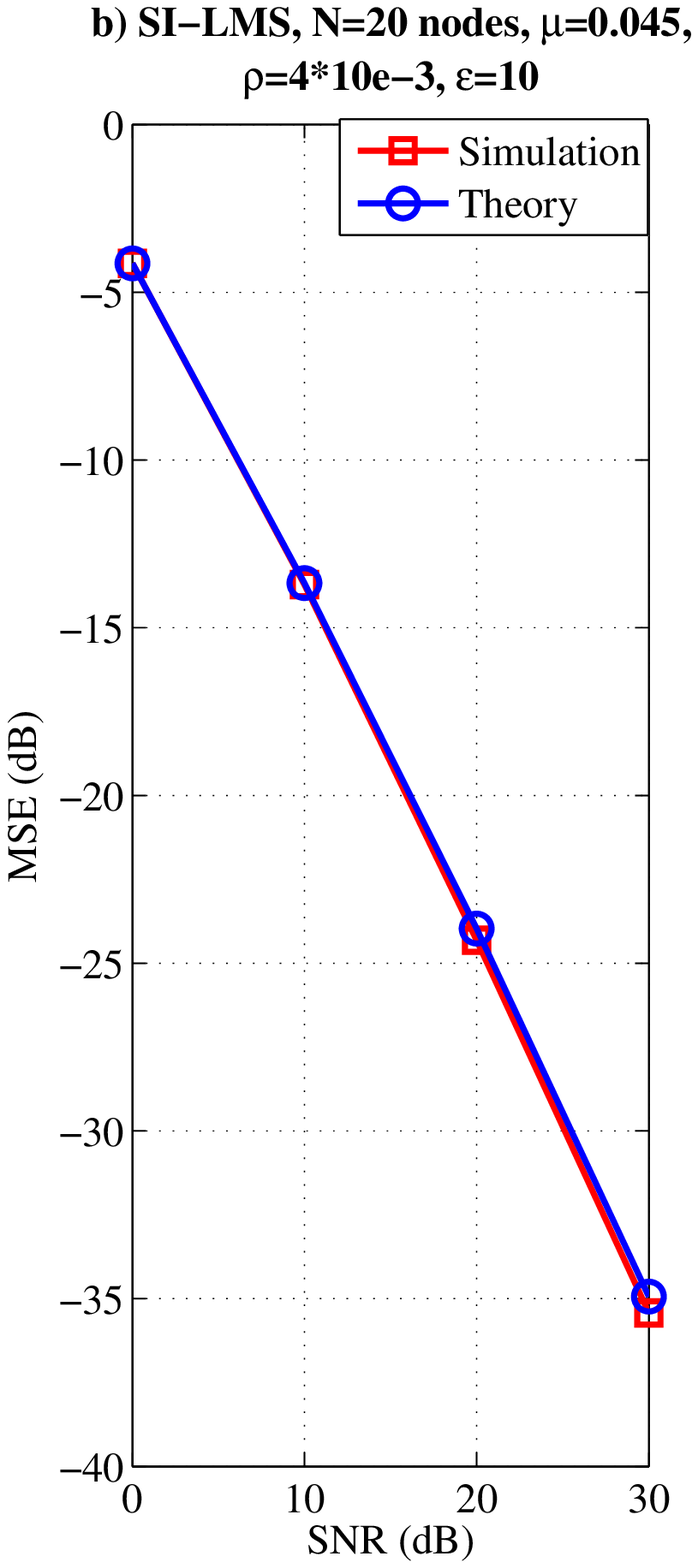}
}\vspace{-1.25em}\caption{\footnotesize MSE performance against SNR
for ES--LMS and SI--LMS}
%\vskip -20pt
\label{fig7}
\end{figure}

\begin{figure}
%\vskip -15pt
\centerline{
\includegraphics[width=0.22\textwidth,height=0.45\textwidth]{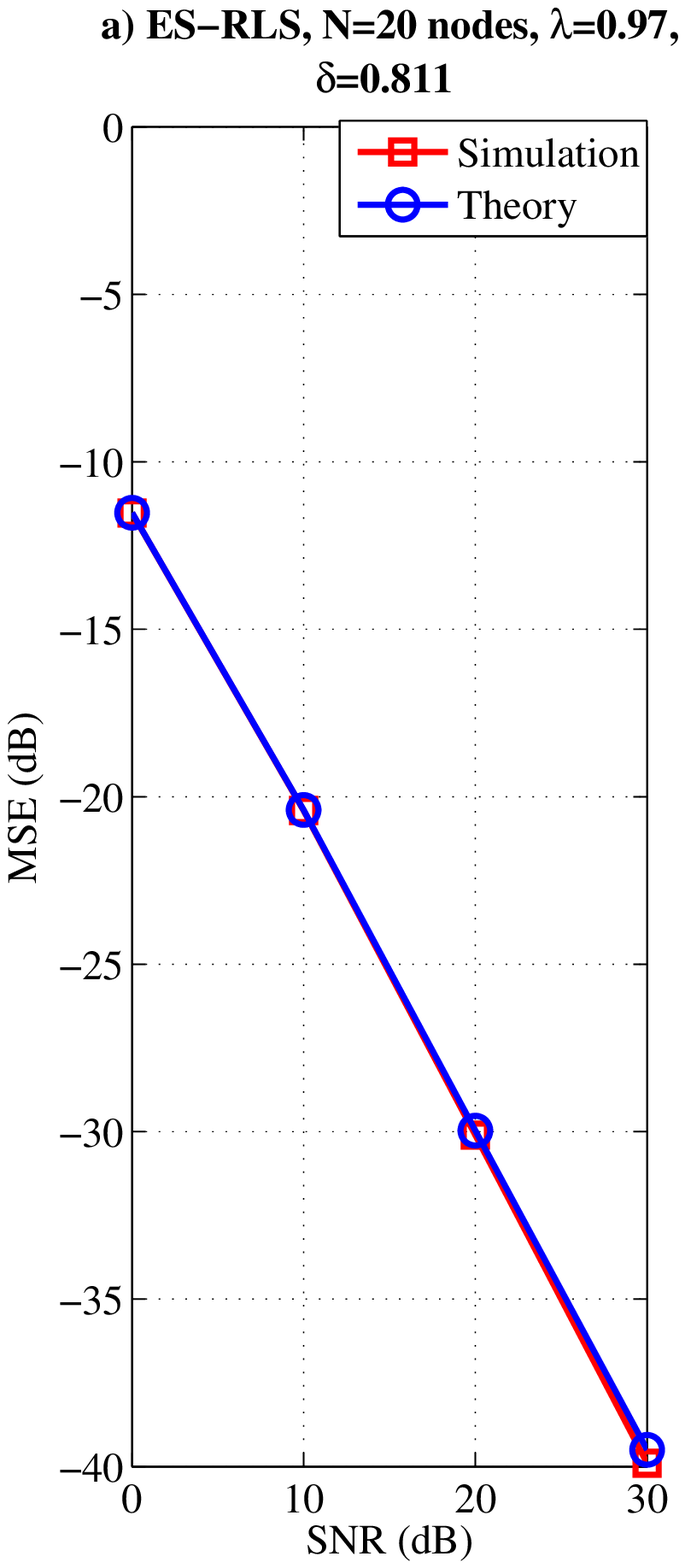}
\includegraphics[width=0.22\textwidth,height=0.45\textwidth]{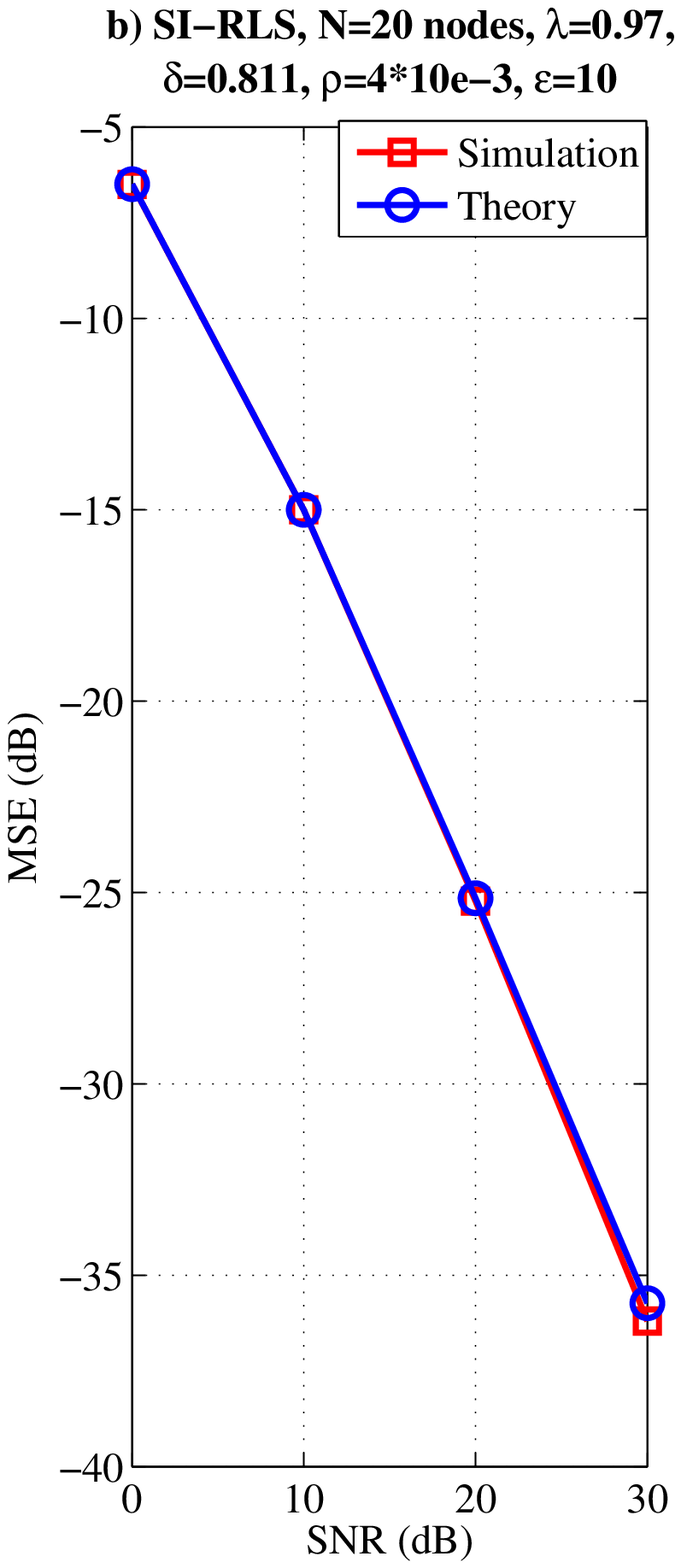}
}\vspace{-1.25em}\caption{\footnotesize MSE performance against SNR
for ES--RLS and SI--RLS} \vskip -10pt \label{fig8}
\end{figure}

\vspace{-0.3em}The aim of this section is to
validate the analytical results obtained in Section IV. First, we
verify the MSE steady--state performance. Specifically, we compare
the analytical results in (\ref{eslms8}), (\ref{silms2}),
(\ref{esrls8}) and (\ref{sirls2}) to the results obtained by
simulations under different SNR values. The SNR indicates the signal
variance to noise variance ratio. We assess the MSE against the SNR,
as show in Figs. \ref{fig7} and \ref{fig8}. For ES--RLS and SI--RLS
algorithms, we use (\ref{esrls8}) and (\ref{sirls2}) to compute the
MSE after convergence. The results show that the analytical curves
coincide with those obtained by simulations and are within 0.01 dB
of each other, which indicates the validity of the analysis. We have
assessed the proposed algorithms with SNR equal to 0dB, 10dB, 20dB
and 30dB, respectively, with 20 nodes in the network. For the other
parameters, we follow the same definitions used to obtain the
network MSE curves in a static scenario. The details have been shown
on the top of each sub figure in Figs. \ref{fig7} and \ref{fig8}.
The theoretical curves for ES--LMS/RLS and SI--LMS/RLS are all close
to the simulation curves and remain within 0.01 dB of one another.

\begin{figure}
\centerline{
\includegraphics[width=0.22\textwidth,height=0.45\textwidth]{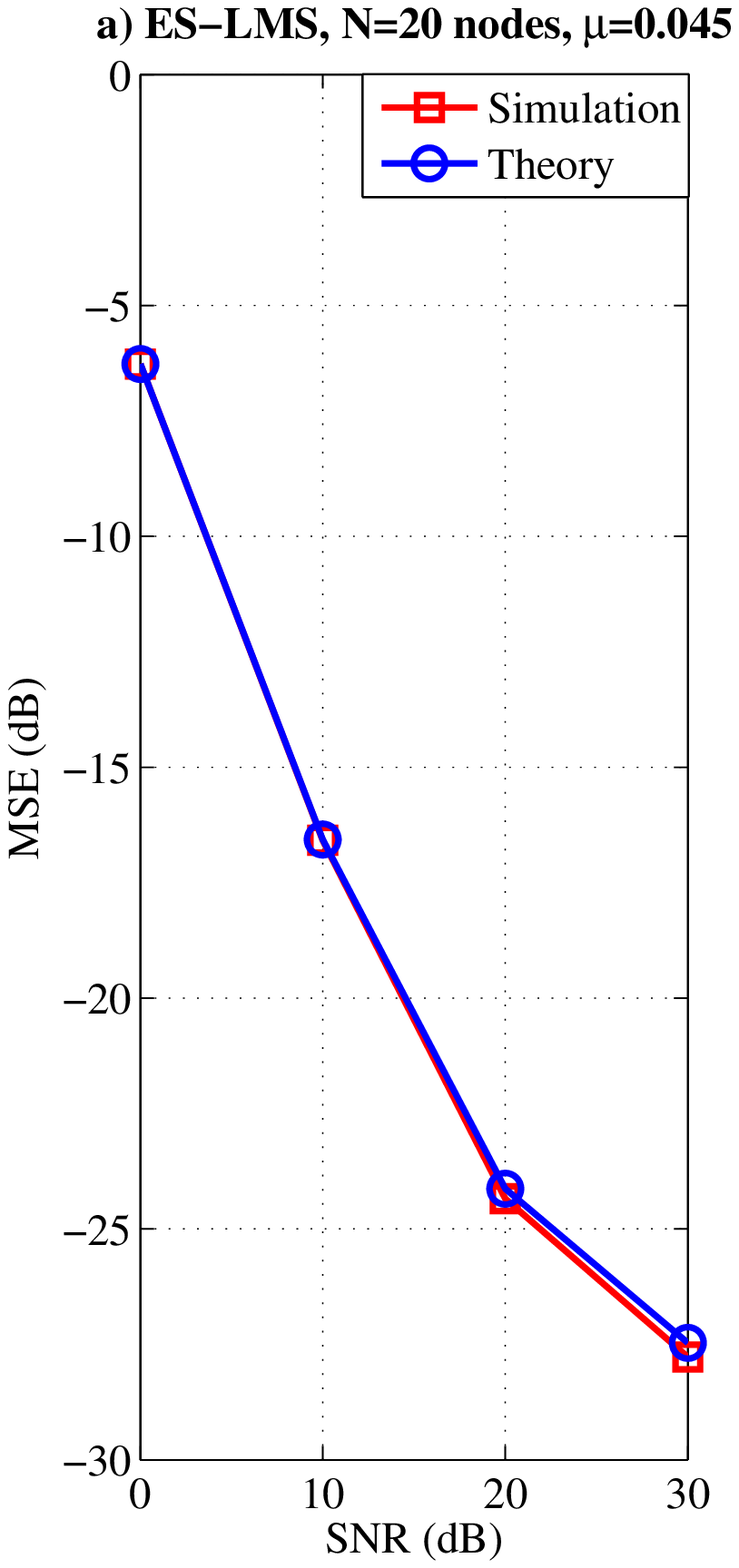}
\includegraphics[width=0.22\textwidth,height=0.45\textwidth]{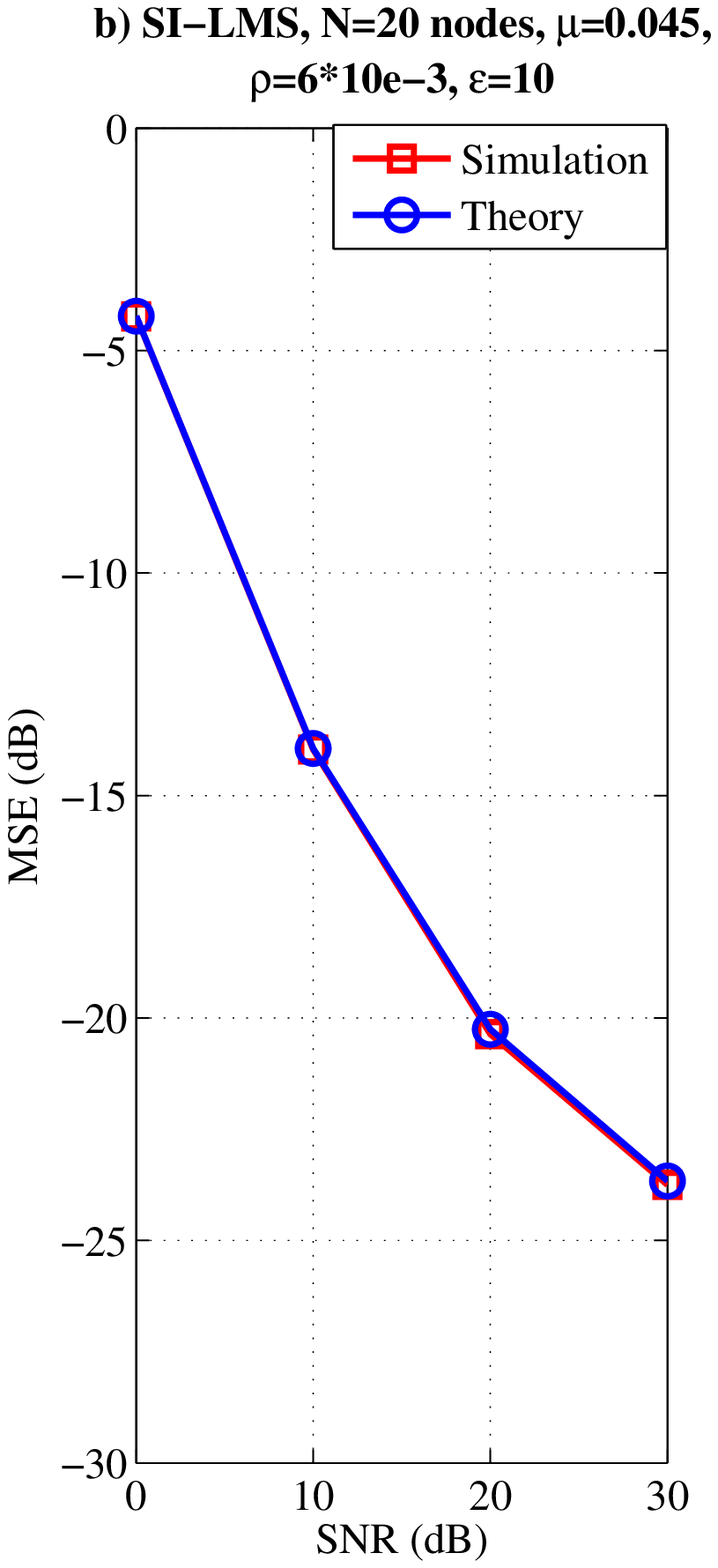}
}\vspace{-1.55em}\caption{\footnotesize MSE performance against SNR
for ES--LMS and SI--LMS in a time–-varying scenario}
%\vskip -20pt
\label{fig9}
\end{figure}

\begin{figure}
%\vskip -15pt
\centerline{
\includegraphics[width=0.22\textwidth,height=0.45\textwidth]{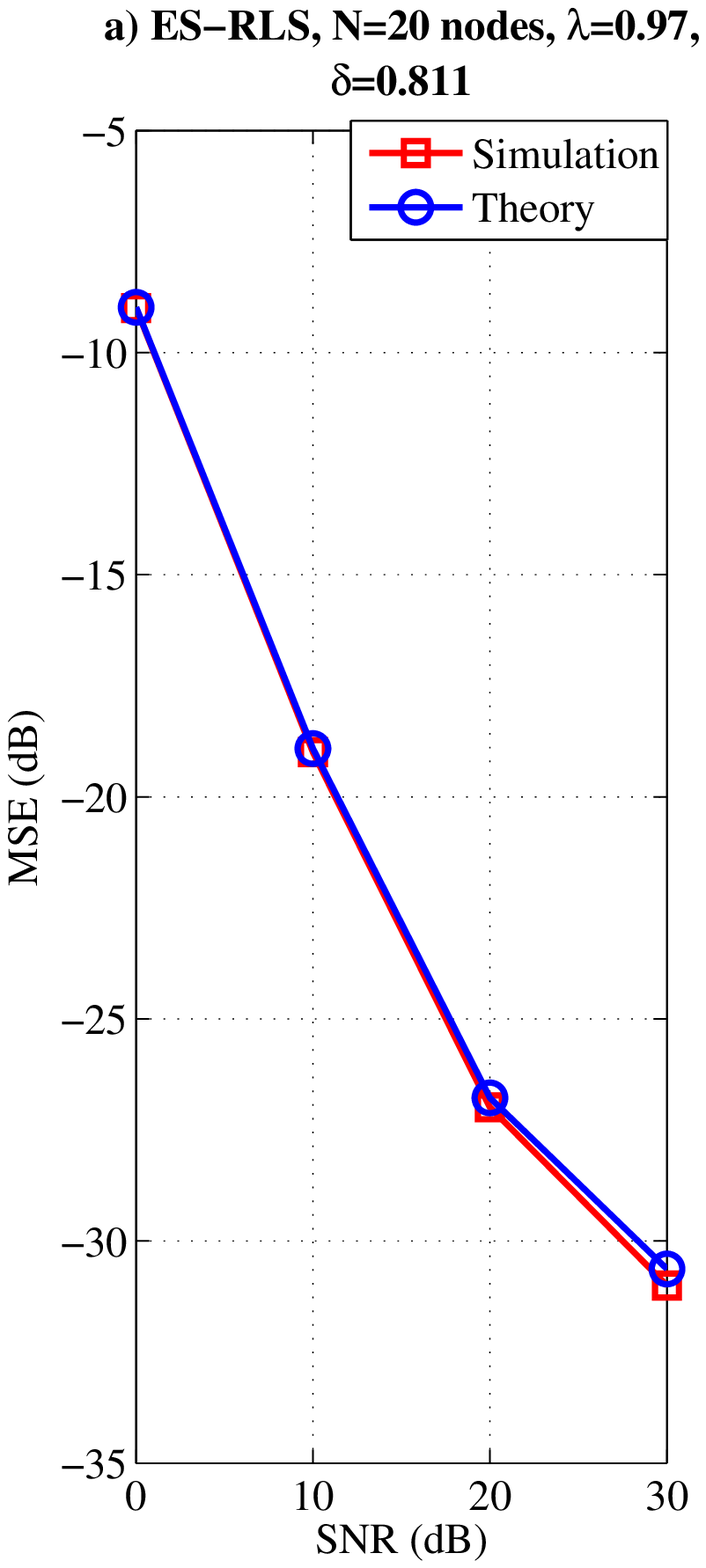}
\includegraphics[width=0.22\textwidth,height=0.45\textwidth]{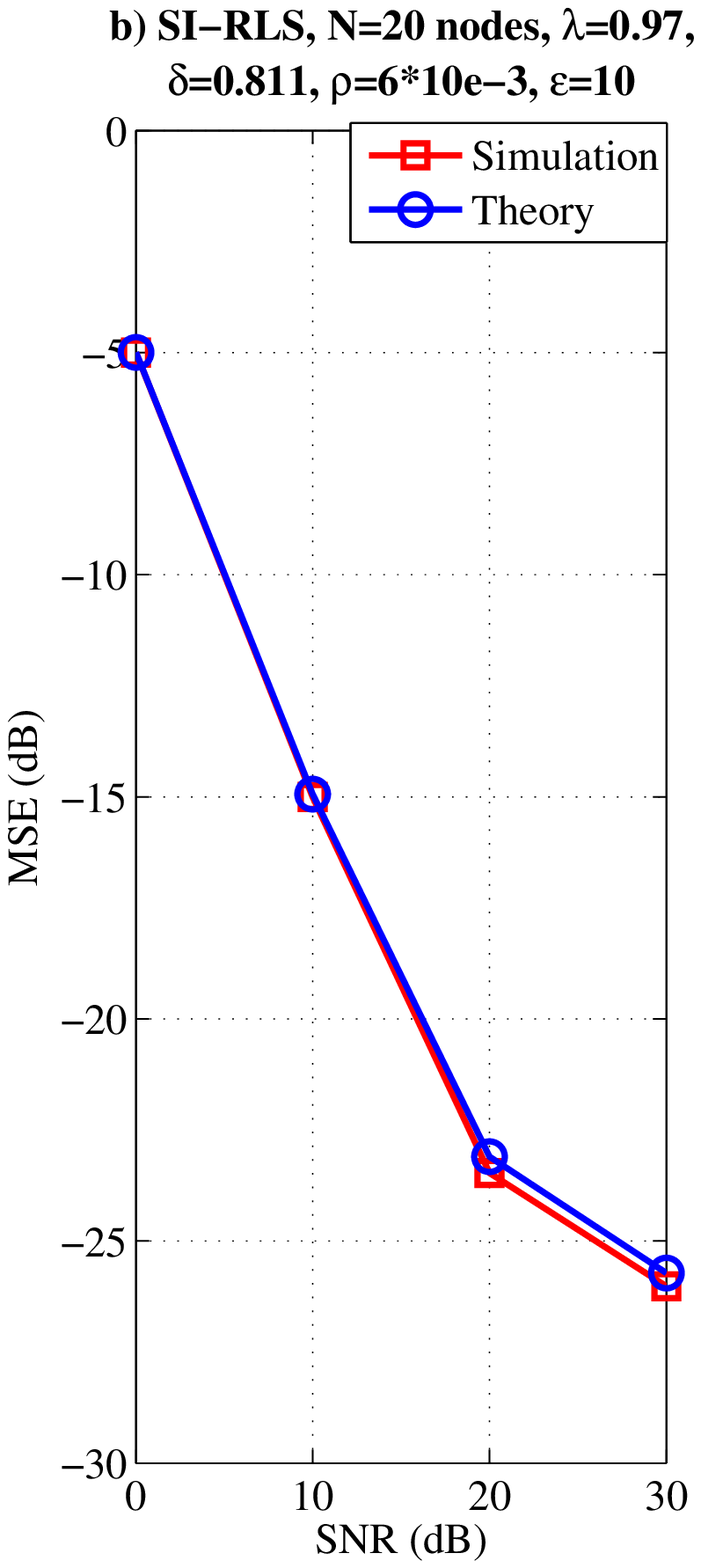}
}\vspace{-1.55em}\caption{\footnotesize MSE performance against SNR
for ES--RLS and SI--RLS in a time–-varying scenario} \vskip -15pt \label{fig10}
\end{figure}

The tracking analysis of the proposed algorithms in
a time--varying scenario is discussed as follows. Here, we verify
that the results (\ref{eslms9}), (\ref{silms3}), (\ref{esrls9}) and
(\ref{sirls3}) of the subsection on the tracking analysis can
provide a means of estimating the MSE. We consider the same model as
in (\ref{tv}). In the next examples, we employ $N=20$ nodes in the
network and the same parameters used to obtain the network MSE
curves in a time--varying scenario. A comparison of the curves
obtained by simulations and by the analytical formulas is shown in
Figs. \ref{fig9} and \ref{fig10}. From these curves, we can verify
that the gap between the simulation and analysis results are within
0.02dB under different SNR values. The details of the parameters are
shown on the top of each sub figure in Figs. \ref{fig9} and
\ref{fig10}.

\subsection{Smart Grids}
The proposed algorithms provide a cost--effective tool that could be used for distributed state estimation in smart grid applications.
In order to test the proposed algorithms in a possible smart grid scenario, we consider the IEEE 14--bus system \cite{Bose}, where 14 is the
number of substations. At every time instant $i$, each bus $k,k=1,2,
\ldots, 14 ,$ takes a scalar measurement $d_k(i)$ according to
\begin{equation}
{d_k(i)}= {X_k \big({\boldsymbol \omega}_0(i)\big)+ n_k(i)},~~~
k=1,2, \ldots, 14 \label{z_k},
\end{equation}
where $\boldsymbol \omega_0(i)$ is the state vector of the entire
interconnected system, $X_k({\boldsymbol \omega}_0(i))$ is a
nonlinear measurement function of bus $k$. The quantity ${n_k(i)}$
is the measurement error with mean equal to zero and which
corresponds to bus $k$.

\begin{figure}[!htb]
\begin{center}
\vskip -10pt
\def\epsfsize#1#2{0.5\columnwidth}
\epsfbox{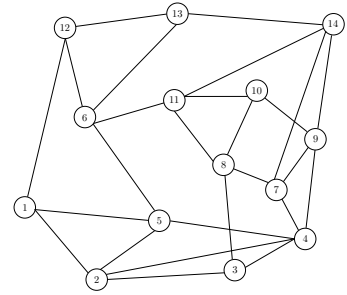} \vspace{-0.5em} \caption{\footnotesize%\vspace{-2.85em}
IEEE 14--bus system for simulation}
\vskip -10pt
\label{fig44}
\end{center}
\end{figure}

Initially, we focus on the linearized DC state estimation problem.
The system is built with 1.0 per unit (p.u) voltage magnitudes at
all buses and j1.0 p.u. branch impedance. Then, the state vector
${\boldsymbol \omega}_0(i)$ is taken as the voltage phase angle
vector ${\boldsymbol \omega}_0$ for all buses. Therefore, the
nonlinear measurement model for state estimation (\ref{z_k}) is
approximated by
\begin{equation}
{d_k(i)}= {\boldsymbol x_k^H(i)\boldsymbol \omega_0+ n_k(i)},~~~
k=1,2, \ldots, 14 ,
\end{equation}
where ${\boldsymbol x}_k(i)$ is the measurement Jacobian vector for
bus $k$. Then, the aim of the distributed estimation algorithm is to
compute an estimate of ${\boldsymbol \omega}_0$, which can minimize
the cost function given by
\begin{equation}
{J_{\boldsymbol\omega_k(i)}({\boldsymbol \omega_k(i)})} =
{\mathbb{E} |{ d_k(i)}- {\boldsymbol x_k^H(i)}{\boldsymbol
\omega_k(i)}|^2} \label{cost function2}.
\end{equation}
We compare the proposed algorithms with the
$\mathcal{M}$--$\mathcal{CSE}$ algorithm \cite{Xie}, the single link
strategy \cite{Zhao}, the standard diffusion RLS algorithm
\cite{Cattivelli2} and the standard diffusion LMS algorithm
\cite{Lopes2} in terms of MSE performance and the Phase Angle Gap.
The MSE comparison is used to determine the accuracy of the
algorithms, and the Phase Angle Gap is used to compare the rate of
convergence. In our scenario, 'Phase Angle Gap' refers to the phase
angle difference between the actual parameter vector or target
${\boldsymbol \omega}_0$ and the estimate $\boldsymbol \omega_k(i)$
for all buses. We define the IEEE--14 bus system as in Fig.
\ref{fig44}.

All buses are corrupted by additive white Gaussian noise with equal
variance $\sigma^2=0.001$. The step size for the standard diffusion
LMS \cite{Lopes2}, the proposed ES--LMS and SI--LMS algorithms is
0.018. The parameter vector $\boldsymbol \omega_0$ is set to an
all--one vector. For the diffusion RLS, ES--RLS and SI--RLS
algorithms the forgetting factor $\lambda$ is set to 0.945 and
$\delta$ is equal to 0.001. The sparsity parameters of the
SI--LMS/RLS algorithms are set to $\rho=0.07$ and $\varepsilon=10$.
The results are averaged over 100 independent runs. We simulate the
proposed algorithms for smart grids under a static scenario.

\begin{figure}[!htb]
\begin{center}
\vskip -10pt
\def\epsfsize#1#2{0.85\columnwidth}
\epsfbox{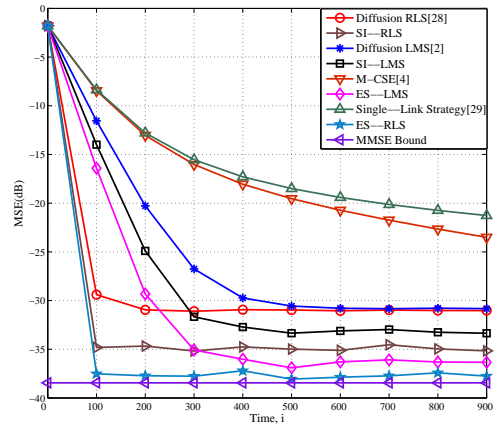} \vspace{-1.25em} \caption{\footnotesize%\vspace{-2.85em}
MSE  performance curves for smart grids.} \vskip -15pt \label{fig55}
\end{center}
\end{figure}

From Fig. \ref{fig55}, it can be seen that ES--RLS has the best
performance, and significantly outperforms the standard diffusion
LMS \cite{Lopes2} and the $\mathcal{M}$--$\mathcal{CSE}$ \cite{Xie}
algorithms. The ES--LMS is slightly worse than ES--RLS, which
outperforms the remaining techniques. SI--RLS is worse than ES--LMS
but is still better than SI--LMS, while SI--LMS remains better than
the diffusion RLS, LMS, $\mathcal{M}$--$\mathcal{CSE}$ algorithms
and the single link strategy.

\begin{figure}[!htb]
\begin{center}
\def\epsfsize#1#2{0.85\columnwidth}
\epsfbox{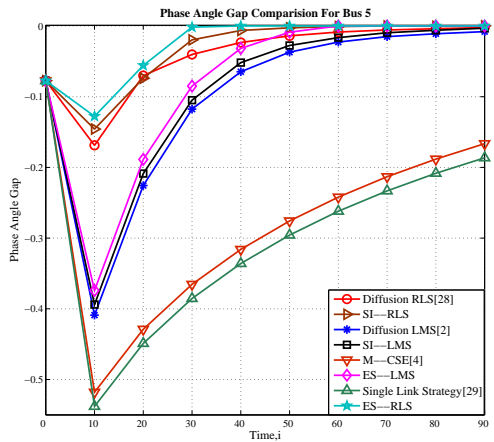} \vspace{-1.25em}
\caption{\footnotesize%\vspace{-2.85em}
Phase Angle Gap comparison for Bus 5.} \vskip -15pt \label{fig66}
\end{center}
\end{figure}

In order to compare the convergence rate, we employ the 'Phase Angle
Gap' to describe the results. We choose bus 5 and the first 90
iterations as an example to illustrate our results. In Fig.
\ref{fig66}, ES--RLS has the fastest convergence rate, while SI--LMS
is the second fastest, followed by the standard diffusion RLS,
ES--LMS, SI--LMS, and the standard diffusion LMS algorithms. The
$\mathcal{M}$--$\mathcal{CSE}$ algorithm and the single link
strategy have the worst performance. The estimates
$\boldsymbol\omega_k(i)$ obtained by the proposed algorithms quickly
reach the target ${\boldsymbol \omega}_0$, which means the Phase
Angle Gap will converge to zero.

\section{Conclusions}

In this paper, we have proposed ES--LMS/RLS and SI--LMS/RLS
algorithms for distributed estimation in applications such as
wireless sensor networks and smart grids. We have compared the
proposed algorithms with existing methods. We have also devised
analytical expressions to predict their MSE steady--state
performance and tracking behavior. Simulation experiments have been
conducted to verify the analytical results and illustrate that the
proposed algorithms significantly outperform the existing
strategies, in both static and time--varying scenarios, in examples
of wireless sensor networks and smart grids.

\section{Acknowledgements}

The authors wish to thank the anonymous reviewers, whose comments
and suggestions have greatly improved the presentation of these
results.

%To briefly
%conclude the updates in the revised manuscript, we made the
%following changes:
%\begin{itemize}
%\item In Section I, the text has been polished and several parts have been rewritten.
%\item Parts of section II have been rewritten and reformulated following the comments of the reviewers, with the description of the combination rules shortened.
%\item The description of the proposed algorithms has been reformulated and now further details about the SI type algorithms are given in section III.
%\item Thanks for the reviewer 2 with the suggestion of parameter $\alpha_{kl}$, the section IV has been completely rewritten.
%\item Section V has been reformulated with new examples and figures.
%\end{itemize}

%\appendices
%\section{A}
%
%%\section*{Acknowledgment}
%%
%%
%%The authors would like to thank...
%%
%%
%%% Can use something like this to put references on a page
%%% by themselves when using endfloat and the captionsoff option.
%%\ifCLASSOPTIONcaptionsoff
%%  \newpage
%%\fi
%
\bibliographystyle{IEEEtran}
\bibliography{reference}
\end{document}